
\input harvmac

\noblackbox
\def\mapright#1{\smash{\mathop{\longrightarrow}\limits^{#1}}}

\lref\BDG{T.~Banks, M.~Dine and M.~Graesser,
``Supersymmetry, axions and cosmology,'' hep-ph/0210256.}

\lref\torsion{A.~Strominger, ``Superstrings with Torsion,'' Nucl.
Phys. {\bf B274} (1986) 253.}

\lref\Choi{K.~Choi, ``Axions and the strong CP problem in
M-theory,'' Phys.\ Rev.\ D {\bf 56} (1997) 6588, hep-th/9706171.}

\lref\BS{S.~M.~Barr and D.~Seckel,
``Planck scale corrections to axion models,''
Phys.\ Rev.\ D {\bf 46}, 539 (1992).}

\lref\MSY{H.~Murayama, H.~Suzuki and T.~Yanagida, ``Radiative
breaking of Peccei-Quinn symmetry at the intermediate mass
scale,'' Phys.\ Lett.\ B {\bf 291} (1992) 418.}

\lref\AMV{M. Atiyah, J. Maldacena and C. Vafa, ``An M-theory
Flop as a Large N Duality,'' J. Math. Phys. {\bf 42} (2001)
3209, hep-th/0011256.}

\lref\shortfl{G. Cardoso, G. Curio, G. Dall'Agata and D. L\"ust,
``Heterotic String Theory on non-Kaehler Manifolds with H-Flux and
Gaugino Condensate,'' hep-th/0310021.}

\lref\BDaxion{T.~Banks and M.~Dine, ``The cosmology of string
theoretic axions,'' Nucl.\ Phys.\ B {\bf 505} (1997) 445,
hep-th/9608197.}

\lref\PQ{R.~D.~Peccei and H.~R.~Quinn,
Phys.\ Rev.\ Lett.\  {\bf 38} 1440 (1977) 1440;
Phys.\ Rev.\ D {\bf 16} (1977) 1791.}

\lref\bbd{K.~Becker, M.~Becker, K.~Dasgupta, P.~Green and E.~
Sharpe, ``Compactifications of Heterotic Strings on Non-K\"ahler
Complex Manifolds: II,'' hep-th/0310058.}

\lref\KM{D.~B.~Kaplan and A.~V.~Manohar, ``Current Mass Ratios Of
The Light Quarks,'' Phys.\ Rev.\ Lett.\  {\bf 56} (1986) 2004.}

\lref\Barr{S.~M.~Barr, ``Solving The Strong CP Problem Without The
Peccei-Quinn Symmetry,'' Phys.\ Rev.\ Lett.\  {\bf 53} (1984)
329.}

\lref\Nelson{A.~E.~Nelson, ``Naturally Weak CP Violation,'' Phys.\
Lett.\ B {\bf 136} (1984) 387.}

\lref\CY{P.~Candelas, G.~Horowitz, A.~Strominger, and E.~Witten,
``Vacuum Configurations for Superstrings,'' Nucl. Phys. {\bf B258}
(1985) 46.}

\lref\SW{A.~Strominger and E.~Witten, ``New Manifolds For
Superstring Compactification,'' Commun.\ Math.\ Phys.\  {\bf 101}
(1985) 341.}

\lref\Polchinski{J.~Polchinski, {\it{String Theory, Volume II:
Superstring Theory and Beyond,}} Cambridge University Press,
1998.}

\lref\Wsb{E.~Witten, ``Symmetry Breaking Patterns In Superstring
Models,'' Nucl.\ Phys.\ B {\bf 258} (1985) 75.}

\lref\Wgtwo{E.~Witten, ``Anomaly Cancellation On Manifolds Of
$G_2$ Holonomy,'' hep-th/0108165.}

\lref\AcharyaYuk{B.~Acharya, ``Compactification with Flux and
Yukawa Hierarchies,'' hep-th/0303234.}

\lref\Wdeconstr{E.~Witten, ``Deconstruction, $G_2$ holonomy, and
Doublet-Triplet Splitting,'' hep-ph/0201018.}

\lref\SYZ{A.~Strominger, S.~T.~Yau, and E.~Zaslow, ``Mirror
Symmetry is T-duality,'' Nucl. Phys. {\bf B479} (1996) 243,
hep-th/9606040.}

\lref\GYZ{S.~Gukov, S.~T.~Yau, and E.~Zaslow, ``Duality and
Fibrations on G(2) manifolds,'' hep-th/0203217.}

\lref\Aspinwall{P.~S.~Aspinwall and C.~A.~Lutken, ``Geometry Of
Mirror Manifolds,'' Nucl.\ Phys.\ B {\bf 353} (1991) 427;
``Quantum Algebraic Geometry Of Superstring Compactifications,''
Nucl.\ Phys.\ B {\bf 355} (1991) 482.}

\lref\triples{J.~de Boer, R.~Dijkgraaf, K.~Hori, A.~Keurentjes,
J.~Morgan, D.~R.~Morrison, and S.~Sethi, ``Triples, fluxes, and
strings,'' Adv.\ Theor.\ Math.\ Phys.\  {\bf 4} (2002) 995,
hep-th/0103170.}

\lref\phases{E.~Witten, ``Phases of $N=2$ Theories in Two
Dimensions,'' Nucl. Phys. {\bf B403} (1993) 159, hep-th/9301042.}

\lref\brg{P.~Aspinwall, B.~Greene and D.~Morrison, ``Calabi-Yau
Moduli Space, Mirror Manifolds, and Space-Time Topology Change in
String Theory,'' Nucl. Phys. {\bf B416} (1994) 414,
hep-th/9309097.}

\lref\sixd{See e.g.: D.~Morrison and C.~Vafa, ``Compactifications
of F-theory on Calabi-Yau Threefolds I,'' Nucl.Phys. {\bf B473}
(1996) 74, hep-th/9602114; ``Compactifications of F-theory on
Calabi-Yau Threefolds II,'' Nucl.Phys. {\bf B476} (1996) 437,
hep-th/9603161\semi N.~Seiberg and E.~Witten, ``Comments on String
Dynamics in Six Dimensions,'' hep-th/9603003.}

\lref\GV{R.~Gopakumar and C.~Vafa, ``Branes and Fundamental
Groups,'' Adv.\ Theor.\ Math.\ Phys.\  {\bf 2} (1998) 399,
hep-th/9712048.}

\lref\Bhet{M.~Becker and D.~Constantin, ``A Note on Flux Induced
Superpotentials in String Theory,'' hep-th/0210131.}

\lref\FW{T.~Friedmann and E.~Witten, ``Unification Scale, Proton
Decay, And Manifolds Of $G_2$ Holonomy,'' hep-th/0211269.}

\lref\hettwo{S.~Kachru and C.~Vafa, ``Exact Results for N=2
Compactifications of Heterotic Strings,'' Nucl. Phys. {\bf B450}
(1995) 69, hep-th/9505105\semi S.~Ferrara, J.~Harvey,
A.~Strominger and C.~Vafa, ``Second-Quantized Mirror Symmetry,''
Phys. Lett. {\bf B361} (1995) 59, hep-th/9505162.}

\lref\KlebanovW{I.~R.~Klebanov and E.~Witten, ``Proton Decay in
Intersecting D-brane Models,'' hep-th/0304079.}

\lref\drsw{M.~Dine, R.~Rohm, N.~Seiberg, and E.~Witten, ``Gluino
Condensation in Superstring Models,'' Phys. Lett. {\bf B156}
(1985) 55. }

\lref\din{J.P.~Derendinger, L. E.~Ibanez, and H.P.~Nilles, ``On
the Low Energy D=4, N=1 Supergravity Theory Extracted from the
D=10, N=1 Superstring,'' Phys. Lett. {\bf B155} (1985) 65.}

\lref\Eva{E.~Silverstein, ``(A)dS Backgrounds from Asymmetric
Orientifolds,'' hep-th/0106209.}

\lref\otherc{C.~Escoda, M.~Gomez-Reino and F.~Quevedo, ``Saltatory
de Sitter String Vacua,'' hep-th/0307160\semi A.~Frey, M.~Lippert
and B.~Williams, ``The Fall of Stringy de Sitter,'' Phys. Rev.
{\bf D68} (2003) 046008, hep-th/0305018.}

\lref\cosmod{
G. Coughlan, W. Fischler, E. Kolb, S. Raby and G. Ross, ``Cosmological
Problems for the Polonyi Potential,'' Phys. Lett. {\bf B131} (1983)
59\semi
B. de Carlos, J.A. Casas, F. Quevedo, and E. Roulet,
``Model Independent Properties and Cosmological Implications of
the Dilaton and Moduli Sectors of 4-D Strings,'' Phys. Lett. {\bf
B318} (1993) 447, hep-ph/9308235\semi
T. Banks, D.B. Kaplan and
A.E. Nelson, ``Cosmological Implications of Dynamical
Supersymmetry Breaking,'' Phys. Rev. {\bf D49} (1994) 779,
hep-ph/9308292.}

\lref\wilczek{J. Feng, J. March-Russell, S. Sethi, and F. Wilczek,
``Saltatory Relaxation of the Cosmological Constant,'' Nucl. Phys.
{\bf B602} (2001) 307, hep-th/0005276.}

\lref\edold{E. Witten, ``New Issues in Manifolds of $SU(3)$
Holonomy,'' Nucl. Phys. {\bf B268} (1986) 79.}

\lref\Vafa{C. Vafa, ``Superstrings and Topological Strings at
Large $N$,'' J. Math. Phys. {\bf 42} (2001) 2798, hep-th/0008142.}

\lref\Friedmann{T. Friedmann, ``On the Quantum Moduli Space of
M-theory Compactifications,'' Nucl. Phys. {\bf B635} (2002) 384,
hep-th/0203256.}

\lref\nonkahler{ K. Dasgupta, G. Rajesh, and S. Sethi, ``M-theory,
Orientifolds and G-flux,'' JHEP {\bf 9908} (1999) 023\semi K.
Becker and K. Dasgupta, ``Heterotic Strings with Torsion,'' JHEP
{\bf 0211} (2002) 006, hep-th/0209077\semi S. Gurrieri, J. Louis,
and A. Micu, ``Mirror Symmetry in Generalized Calabi-Yau
Compactifications,'' Nucl. Phys. {\bf B654} (2003) 61,
hep-th/0211102\semi G.L. Cardoso, G. Curio, G. Dall'Agata, D.
L\"ust, P. Manousselis, and G. Zoupanos, ``Non-Kaehler String
Backgrounds and their Five Torsion Classes,'' Nucl. Phys. {\bf
B652} (2003) 5, hep-th/0211118\semi S. Kachru, M. Schulz, P.
Tripathy, and S. Trivedi, ``New Supersymmetric String
Compactifications,'' JHEP {\bf 0303} (2003) 061,
hep-th/0211182\semi K. Becker, M. Becker, K. Dasgupta, and P.
Green, ``Compactifications of Heterotic Theory on Non-K\"ahler
Complex Manifolds I, '' JHEP {\bf 0304} (2003) 007,
hep-th/0301161.}

\lref\bbdp{ K. Becker, M. Becker, K. Dasgupta, and S. Prokushkin,
``Properties of Heterotic Vacua from Superpotentials,''
hep-th/0304001.}

\lref\casas{B. de Carlos, J.A. Casas, and C. Munoz,
``Supersymmetry Breaking and Determination of the Unification
Gauge Coupling Constants in String Theories,'' Nucl. Phys. {\bf
B399} (1993) 623, hep-th/9204012.}

\lref\boundtwo{G. Curio and A. Krause, ``Four-flux and Warped Heterotic
M-theory Compactifications,'' Nucl. Phys. {\bf B602} (2001) 172,
hep-th/0012152.}

\lref\schimmrigk{R. Schimmrigk, ``Scaling Behavior in String
Theory,'' Phys. Lett. {\bf B388} (1996) 60, hep-th/9412077.}

\lref\threegen{See e.g.: B.R. Greene, K. Kirklin, P. Miron and G.G. Ross,
``A Three Generation Superstring Model I: Compactification and
Discrete Symmetries,'' Nucl. Phys. {\bf B278} (1986) 667\semi
D. Gepner, ``String Theory on Calabi-Yau Manifolds: The Three
Generations Case,'' hep-th/9301089\semi
S. Kachru, ``Some Three Generation (0,2) Calabi-Yau Models,''
Phys. Lett. {\bf B349} (1995) 76, hep-th/9501131\semi
B. Ovrut, T. Pantev and R. Reinbacher, ``Torus Fibered Calabi-Yau
Threefolds with Nontrivial Fundamental Group,'' JHEP {\bf 0305}
(2003) 040, hep-th/0212221.}

\lref\ovrut{ G. Moore, G. Peradze, and N. Saulina, ``Instabilities
in Heterotic M-theory Induced by Open Membrane Instantons,'' Nucl.
Phys. {\bf B607} (2001) 117, hep-th/0012104\semi E. Buchbinder, R.
Donagi, and B. Ovrut, ``Superpotentials for Vector Bundle
Moduli,'' Nucl. Phys. {\bf B653} (2003) 400, hep-th/0205190\semi
G. Curio and A. Krause, ``G-Fluxes and Nonperturbative
Stabilisation of Heterotic M-theory,'' Nucl. Phys. {\bf B643}
(2002) 131, hep-th/0108220.}

\lref\rohmw{R. Rohm and E. Witten, ``The Antisymmetric Tensor
Field in Superstring Theory,'' Annals Phys. {\bf 170} (1986) 454.}

\lref\anomaly{E. Witten, ``Global Anomalies in String Theory,'' in
Symposium on Anomalies, Geometry, Topology; Bardeen and White,
eds., World Scientific, 1985.}

\lref\worldsheet{E. Witten,``World-Sheet Corrections Via
D-Instantons,'' JHEP {\bf 0002} (2000) 030, hep-th/9907041.}

 \lref\sethi{A. Keurentjes and S. Sethi,
``Twisting $E_8$ Five-Branes,'' Phys. Rev. {\bf D66} (2002)
046001, hep-th/0205162.}

\lref\douglas{M. Douglas, ``The Statistics of String / M-theory
Vacua,'' hep-th/0303194\semi
S. Ashok and M. Douglas, ``Counting Flux Vacua,'' hep-th/0307049.}

\lref\seiberg{
N. Seiberg, ``Exact Results on the Space of Vacua of Four Dimensional
SUSY Gauge Theories,'' Phys. Rev. {\bf D49} (1994) 6857\semi
K.Intriligator, R.G.Leigh, and N. Seiberg, ``Exact
Superpotentials in Four Dimensions,'' Phys. Rev. {\bf D50} (1994)
1092, hep-th/9403198}

\lref\pouliot{D.Finnell and P.Pouliot, ``Instanton Calculations
versus Exact Results in 4 Dimensional SUSY Gauge Theories,''
hep-th/9503115.}

\lref\susskind{L. Susskind, ``The Anthropic Landscape of String
Theory,'' hep-th/0302219.}

\lref\BP{R. Bousso and J. Polchinski, ``Quantization of Four-Form
Fluxes and Dynamical Neutralization of the Cosmological Constant,''
JHEP {\bf 0006} (2000) 006, hep-th/0004134.}

\lref\kklt{S. Kachru, R. Kallosh, A. Linde, and S. Trivedi, ``de
Sitter Vacua in String Theory,'' Phys. Rev. {\bf D68} (2003)
046005, hep-th/0301240.}

\lref\bkq{C.P. Burgess, R. Kallosh, and F. Quevedo, ``de Sitter
String Vacua from Supersymmetric D-terms,'' hep-th/0309187.}

\lref\wsinst{E. Silverstein and E. Witten, ``Criteria for
Conformal Invariance of (0,2) Models,'' Nucl. Phys. {\bf B444}
(1995) 161, hep-th/9503212\semi A. Basu and S. Sethi,
``World-sheet Stability of (0,2) Linear Sigma Models,''
hep-th/0303066\semi C. Beasley and E. Witten, ``Residues and
World-Sheet Instantons,'' hep-th/0304115.}

\lref\berg{E. Bergshoeff, M. de Roo, B. de Wit, and P. van
Nieuwenhuizen, ``Ten-Dimensional Maxwell-Einstein Supergravity,
its Currents, and the Issue of its Auxiliary Fields,'' Nucl.Phys.
{\bf B195} (1982) 97.}

\lref\shifv{M.A. Shifman and A.I. Vainshtein, ``On Gluino
Condensation in Supersymmetric Gauge Theories with SU(N) and O(N)
Groups,'' Nucl.Phys. {\bf B296} (1988) 445. }

\lref\horava{P. Ho\v{r}ava, ``Gluino Condensation in Strongly
Coupled Heterotic String Theory,'' Phys.Rev. {\bf D54} (1996)
7561, hep-th/9608019.}

\lref\rxiu{R. Xiu, ``Supersymmetry Breaking Scheme and The
Derivation of $M_{GUT}=10^{16}GeV$ from A String Model,''
hep-ph/9412262.}

\lref\nienil{A. Niemeyer and H.P. Nilles, ``Gaugino Condensation
and the Vacuum Expectation Value of the Dilaton,''
hep-th/9508173.}

\lref\nilles{H.P. Nilles, ``Dynamical Gauge Coupling Constants,''
hep-ph/9601241.}

\lref\nishi{H. Nishi, ``$SU(n)$-Chern-Simons Invariants of Seifert
Fibered 3-Manifolds,'' Int. J. Math {\bf 9} (1998) 295.}

\lref\auckly{D.~Auckly, ``Topological Methods to Compute
Chern-Simons Invariants,'' Math. Proc. Camb. Phil. Soc. {\bf 115}
(1994) 229.}

\lref\conrad{J.O. Conrad, ``On Fractional Instanton Numbers in Six
Dimensional Heterotic E8 x E8 Orbifolds,''  JHEP {\bf 0011 }
(2000) 022, hep-th/0009251.}

\lref\shifvtwo{M.A. Shifman and A.I. Vainshtein, ``On Holomorphic
Dependence and Infrared Effects in Supersymmetric Gauge
Theories,''  Nucl.Phys. {\bf B359 } (1991) 571.}

\lref\bdc{T. Barreiro, B. de Carlos, and E.J. Copeland,
``Stabilizing the Dilaton in Superstring Cosmology,'' Phys.Rev.
 {\bf D58 } (1998) 083513, hep-th/9805005.}

\lref\bgw{P. Bin\'{e}truy, M.K. Gaillard, and Y.-Y. Wu, ``Dilaton
Stabilization in the Context of Dynamical Supersymmetry Breaking
through Gaugino Condensation,'' Nucl.Phys. {\bf B481} (1996) 109,
hep-th/9605170.}

\lref\yyw{Y.-Y. Wu, ``Dilaton Stabilization and Supersymmetry
Breaking by Dynamical Gaugino Condensation in the Linear Multiplet
Formalism of String Effective Theory,'' Nucl.Phys. {\bf B481}
(1996) 109, hep-th/9610089.}

\lref\lk{V. Kaplunovsky and J. Louis, ``Field Dependent Gauge
Couplings in Locally Supersymmetric Effective Quantum Field
Theories,'' Nucl.Phys. {\bf B422} (1994) 57, hep-th/9402005.}

\lref\BCOV{M. Bershadsky, S. Cecotti, H. Ooguri, and C. Vafa,
``Kodaira-Spencer Theory of Gravity and Exact Results for Quantum
String Amplitudes,'' hep-th/9309140.}

\lref\KL{V.Kaplunovsky and J.Louis, ``On Gauge Couplings in String
Theory,'' hep-th/9502077.}

\lref\Ov{B. Ovrut,``$N=1$ Supersymmetric Vacua in Heterotic
M-Theory,'' hep-th/9905115.}

\lref\St{S. Stieberger, ``(0,2) Heterotic Gauge Couplings and
their M-Theory Origin,'' hep-th/9807124. }

\lref\GKP{S. Giddings, S. Kachru, and J. Polchinski, ``Hierarchies
from Fluxes in String Compactifications,'' Phys. Rev. {\bf D66}
(2002) 106006, hep-th/0105097.}

\lref\BG{K.~Behrndt and S.~Gukov, ``Domain Walls and
Superpotentials from M-theory on Calabi-Yau Three-Folds,'' Nucl.
Phys. {\bf B580} (2000) 225, hep-th/0001082.}

\lref\TV{T.~R.~Taylor and C.~Vafa, ``RR flux on Calabi-Yau and
partial supersymmetry breaking,'' Phys.\ Lett.\  {\bf B474}(2000)
130, hep-th/9912152.}

\lref\calibr{S.~Gukov, ``Solitons, Superpotentials and Calibrations",
Nucl.\ Phys.\ {\bf B574} (2000) 169, hep-th/9911011.}

\lref\GVW{S.~Gukov, C.~Vafa, and E.~Witten, ``CFT's From
Calabi-Yau Four-folds,'' Nucl.\ Phys.\ {\bf B584} (2000) 69,
hep-th/9906070.}

\lref\mclean{R.~McLean, ``Deformations of Calibrated
Submanifolds,'' Comm. Anal. Geom. {\bf 6} (1998) 705.}

\lref\Rozanskyseif{L.~Rozansky, ``A Large k Asymptotics of Witten's
Invariant of Seifert Manifolds,'' Commun. Math. Phys. {\bf 171} (1995) 279.}

\lref\Wcy{E. Witten, ``Strong Coupling Expansion of
Calabi-Yau Compactification'', hep-th/9602070.}

\lref\LOW{A. Lukas, B. Ovrut, and D. Waldram, ``On the
Four-Dimensional Effective Action of Strongly Coupled Heterotic
String Theory,'' hep-th/9710208.}

\lref\Bobby{B.~S.~Acharya, ``A Moduli
Fixing Mechanism in M theory,'' hep-th/0212294.}

\lref\AW{M. Atiyah and E. Witten, ``M-Theory
Dynamics On A Manifold Of $G_2$ Holonomy,'' hep-th/0107177.}

\lref\AcharyaW{B.~Acharya and E.~Witten, ``Chiral Fermions from
Manifolds of $G_2$ Holonomy,'' hep-th/0109152.}

\lref\Oz{H.~Ita, Y.~Oz, and T.~Sakai, ``Comments on M Theory
Dynamics on G2 Holonomy Manifolds,'' JHEP {\bf 0204} (2002) 001,
hep-th/0203052.}

\lref\Austin{D.M.~Austin, ``$SO(3)$-Instantons on $L(p,q) \times R$,''
J. Diff. Geom. {\bf 32} (1990) 383.}

\lref\FS{R.~Fintushel and R.~Stern, ``Invariants for Homology
3-Spheres,'' London Math. Soc. Lect. Notes {\bf 150} (1989) 125.}

\lref\FL{R.~Fintushel and T.~Lawson, ``Compactness of the Moduli
Spaces for Orbifold Instantons,'' Topology Appl. {\bf 23} (1986)
305.}

\lref\babupati{K.S. Babu and J.C. Pati, ``The problems of
unification-mismatch and low $\alpha_3$: A solution with light
vector-like matter,'' Phys. Lett. {\bf B384} (1996) 140,
hep-ph/9606215.}

\lref\louis{  J. Louis, J. Sonnenschein, S. Theisen, and S.
Yankielowicz, ``Non-Perturbative Properties of Heterotic String
Vacua Compactified on ${K3\times T^2}$,'' Nucl.Phys. {\bf B480}
(1996) 185, hep-th/9606049.}

\lref\bo{ E.I. Buchbinder and B.A. Ovrut, ``Vacuum Stability in
Heterotic M-Theory,'' hep-th/0310112.}

\lref\CveticP{M.~Cvetic and I.~Papadimitriou, ``Conformal Field
Theory Couplings for Intersecting D-branes on Orientifolds,''
Phys.\ Rev.\ D {\bf 68} (2003) 046001, hep-th/0303083}

\lref\CveticSU{M.~Cvetic, G.~Shiu and A.~M.~Uranga, ``Chiral
Four-Dimensional N = 1 Supersymmetric Type IIA Orientifolds from
Intersecting D6-branes,'' Nucl.\ Phys.\ B {\bf 615} (2001) 3,
hep-th/0107166.}

\lref\CveticLW{M.~Cvetic, P.~Langacker and J.~Wang, ``Dynamical
Supersymmetry Breaking in Standard-like Models with Intersecting
D6-branes,'' Phys.\ Rev.\ D {\bf 68} (2003) 046002,
hep-th/0303208.}

\lref\Cardoso{G.~L.~Cardoso, G.~Curio, G.~Dall'Agata and D.~Lust,
``BPS Action and Superpotential for Heterotic String
Compactifications with Fluxes,'' JHEP {\bf 0310} (2003) 004,
hep-th/0306088.}

\font\cmss=cmss10 \font\cmsss=cmss10 at 7pt

\def\IC{\relax\hbox{$\inbar\kern-.3em{\rm C}$}}
\def\IR{\relax{\rm I\kern-.18em R}}
\def\Z{\relax\ifmmode\mathchoice
{\hbox{\cmss Z\kern-.4em Z}}{\hbox{\cmss Z\kern-.4em Z}}
{\lower.9pt\hbox{\cmsss Z\kern-.4em Z}}
{\lower1.2pt\hbox{\cmsss Z\kern-.4em Z}}\else{\cmss Z\kern-.4em
Z}\fi}

\hskip 1cm

\vskip -1 cm

\Title{\vbox{\baselineskip12pt \hbox{hep-th/0310159}
\hbox{HUTP-03/A072}
\hbox{SLAC-PUB-9863}
\hbox{SU-ITP-03/28}}}
{\vbox{
\centerline{Heterotic Moduli Stabilization}
\medskip
{\vbox{\centerline{with Fractional Chern-Simons Invariants }}}}}

\centerline{Sergei
Gukov,$^{a}$\footnote{$^1$}{gukov@tomonaga.harvard.edu} Shamit
Kachru,$^{b,c}$\footnote{$^2$}{skachru@stanford.edu} Xiao
Liu,$^{b,c}$\footnote{$^3$}{liuxiao@itp.stanford.edu} and Liam
McAllister$^{b}$\footnote{$^4$}{lpm@itp.stanford.edu} }
\medskip\centerline{$^a$ \it Department of Physics, Harvard University, Cambridge, MA 02138}
\medskip\centerline{$^b$ \it Department of Physics, Stanford University, Stanford, CA
94305}
\medskip\centerline{$^c$ \it SLAC, Stanford University,  Menlo Park, CA
94309}

\vskip .2in \noindent\

We show that fractional flux from Wilson lines can stabilize the moduli of
heterotic string compactifications on Calabi-Yau threefolds.  We observe
that the Wilson lines used in GUT symmetry breaking naturally induce a
fractional flux.  When combined with a hidden-sector gaugino condensate,
this generates a potential for the complex structure moduli, K\"ahler
moduli, and dilaton. This potential has a supersymmetric AdS minimum at
moderately weak coupling and large volume.  Notably, the necessary
ingredients for this construction are often present in realistic models.
We explore the type IIA dual phenomenon, which involves Wilson lines in
D6-branes wrapping a three-cycle in a Calabi-Yau, and comment on the
nature of the fractional instantons which change the Chern-Simons
invariant.

%
%

\smallskip\
\newsec{Introduction}

When string theory is compactified on a Calabi-Yau manifold \CY,
the resulting low-energy field theory typically contains some
number of massless scalar fields, or moduli.
Gravitational experiments and the requirement
of consistency with nucleosynthesis place rather strong constraints
on the existence of such fields (see e.g. \cosmod).
If moduli were an essential
feature of all string compactifications then model building
would be very difficult.
Fortunately, moduli are only endemic in the simplest, most
symmetric constructions.
General backgrounds involving fluxes, as well as nonperturbative
effects, tend to create
potentials for some or all moduli.  Even so, although
compactifications with reduced moduli spaces are easy to
construct, it remains challenging to eliminate all of the moduli
in a given model.

Two fields which have proven particularly difficult to stabilize
are the Calabi-Yau volume and, in heterotic compactifications, the
dilaton.  The problem is especially acute in these cases because
the dilaton and volume directly influence the gauge and
gravitational couplings in our world, making rolling values
unacceptable.  Moreover, as these parameters govern the string and
sigma-model perturbation expansions, a controllable
compactification requires that the dilaton and volume be
stabilized at weak coupling and large radius.

We will demonstrate that this can be achieved in a certain class
of heterotic compactifications on Calabi-Yau spaces with a large
fundamental group. The context for this proposal is the original
work \drsw\ of Dine, Rohm, Seiberg, and Witten, who observed that
the combination of a gaugino condensate (in the hidden-sector of
the $E_8 \times E_8$ heterotic string) and a background three-form
flux generates a potential for the dilaton but leaves the
cosmological constant zero at tree level.\foot{Closely related
simultaneous work appears in \din.} As was understood there and in
more detail in subsequent work, because of the quantization
condition for the three-form of the heterotic theory, the dilaton
cannot be fixed at weak coupling. The essential difficulty is that
the gaugino condensate term is nonperturbatively small when the
coupling is weak, whereas quantization forces the flux term to be
of order one. The resulting potential drives the dilaton to strong
coupling.
\footline={\hss\tenrm\folio\hss}

It is important, however, that the Chern-Simons contribution to
the heterotic three-form flux does not obey the same
quantization condition as the contribution from the field
strength of the antisymmetric tensor.
In fact, as we will explain below, the Chern-Simons contribution
of a flat gauge bundle can take fractional values of order $1/N$,
where $N$ is related to the order of the fundamental group. On
Calabi-Yau manifolds with sufficiently large fundamental group
this provides a natural mechanism to stabilize the dilaton at weak
coupling.  The same effect stabilizes all K\"ahler moduli once the
dependence of the gauge coupling on these moduli is correctly
incorporated.  For related earlier work see
\refs{\rohmw,\nienil,\nilles,\bgw,\bdc,\yyw,\rxiu}.

The requirements that the Calabi-Yau manifold should have
non-trivial fundamental group and that the gauge bundle should
have nonzero Wilson lines are actually well motivated by other
model-building considerations. In fact, most models of particle
physics based on Calabi-Yau compactifications of the heterotic
string involve manifolds with non-trivial fundamental group and
associated gauge bundles with Wilson lines.

A standard way to construct such manifolds is to quotient a
simply-connected Calabi-Yau space by a freely-acting discrete
symmetry group $G$.  The resulting string GUT model solves a
number of important problems.  For instance, in simple
constructions the number of generations is divided by $\vert G
\vert$, leading to models with realistically low numbers of
generations \CY. Moreover, one can naturally solve the
doublet-triplet splitting problem \refs{\Wsb,\Wdeconstr} in this
setting.

More importantly, the non-trivial fundamental group allows us to
introduce Wilson lines.  In addition to being an attractive method
of GUT symmetry breaking, Wilson lines are actually indispensable,
as standard heterotic string models do not admit adjoint Higgses
of the GUT group \Wsb.

We will add the stabilization of moduli to this list of problems
which admit natural solutions on Calabi-Yau manifolds with
non-trivial fundamental group and non-trivial gauge connection.
The dilaton, K\"ahler moduli, and complex structure moduli can all
be stabilized by incorporating the effects of gaugino condensation
and the flux induced by the Wilson lines.

We would like to underscore the happy coincidence that the
necessary ingredients for our construction are automatically
present in certain realistic models.  Wilson lines typically lead
to Chern-Simons flux, as we will explain in \S3.3.  Thus,
heterotic string GUT models with Wilson-line symmetry breaking
often have a background flux and an associated constant term in
the superpotential. To the best of our knowledge the consequences
of this term have not been well explored in the literature. In a
restricted subset of models, namely those with hidden-sector
gaugino condensation and very small Chern-Simons flux, the effect
is dramatic: the moduli can be fixed, in a controllable regime, by
the mechanism we are proposing.

The organization of this paper is as follows. In \S2 we review
basic facts about the relevant supergravity Lagrangians in ten and
four dimensions, and about the superpotential generated by gaugino
condensation in the hidden $E_8$.  In \S3 we review the
quantization conditions on three-form flux and describe how
fractional flux can arise in the presence of flat connections with
fractional Chern-Simons invariant.  In \S4 we describe how the
fractional flux of \S3 can be combined with gaugino condensation
to stabilize the dilaton at weak coupling, along with the complex
structure moduli.  In \S5 we include loop corrections and show
that it becomes possible to simultaneously stabilize the K\"ahler
moduli as well as the dilaton; this requires more restrictive
assumptions about the choices of gauge bundles.  We observe that a
strong coupling transition naturally arises in this setting and we
provide a toy model which illustrates the smoothness of this
transition.  In \S6 we discuss some basic aspects of the dual
descriptions of our story, including the dual type IIA theories
with wrapped D6-branes.  In \S7 we explore the nature of the
domain walls which interpolate between configurations with
distinct fractional Chern-Simons invariants. We conclude with a
discussion of possible extensions and broader issues in \S8.

As this paper was being finalized, three papers which have some
overlap with our results appeared \refs{\shortfl,\bbd,\bo}.


\newsec{Gaugino Condensation in the Heterotic String}

In this section we review the structure of the heterotic string
low-energy effective Lagrangian, with particular attention to
terms coupling the heterotic three-form flux, $H$, to the
gauginos. In \S2.1 we fix notation by presenting the low-energy
action for the heterotic string in ten dimensions.  We
dimensionally reduce this action on a Calabi-Yau threefold and
describe the potential appearance of a gaugino condensate in the
resulting ${\cal{N}} = 1,d=4$ configuration.  In \S2.2 we show how
to derive the four-dimensional action of \S2.1 from a simple
superpotential induced by the flux and the gaugino condensate. In
\S2.3 we explain that the dilaton potential does not have a
minimum at finite coupling unless the background flux is
fractional.

\subsec{Effective Lagrangian for the Heterotic Theory}

The low-energy effective action for the heterotic string in
ten-dimensional Einstein frame is \berg\ \eqn\tenact{\eqalign{ S =
& {{1\over{2\alpha'^4}}\int d^{10}x
\sqrt{-g_{10}}\left({{{\cal{R}}_{10}
-{{1\over{2}}{\partial_{A}\phi \partial^{A}\phi}}
-{1\over{12}}e^{-\phi}\left({{H_{ABC}
-{{\alpha'}\over{16}}e^{\phi\over{2}}\overline{\chi}_{10}
\Gamma_{ABC}\chi_{10}}}\right)^2}
 }\right.} \cr & \left.
-{\alpha'\over{4}}{e^{-{\phi\over{2}}}{\rm{tr}}(F_{AB}F^{AB})}
-{\alpha'}{\rm{tr}}(\overline{\chi}_{10}
\Gamma^{A}D_{A}\chi_{10})\right)}} Indices $A,B$ run from $0...9$,
and $\mu,\nu$ are four-dimensional spacetime indices.  The
internal space has real indices $m,n$ and (anti)holomorphic
indices $i,j,\bar{i},\bar{j}$. The Einstein-frame metric $g_{10}$
has Ricci scalar ${\cal{R}}_{10}$, while $\omega$ is the spin
connection and $\phi$ is the dilaton. The heterotic string has
gauge field strength $F_{\mu\nu}$ and gaugino field $\chi_{10}$;
all traces are taken in the fundamental representation. The
three-form flux $H_{ABC}$ is defined by
\eqn\hdef{H=dB-{\alpha'\over{4}}\Bigl({
{\Omega_{3}(A)-\Omega_{3}(\omega)}}\Bigr)} where $\Omega_{3}$ is
the Chern-Simons three-form, \eqn\omis{ \Omega_{3}(A) \equiv \tr
\left(A \wedge dA + {2\over{3}}A\wedge A \wedge A \right)} with a
similar formula for $ \Omega_{3}(\omega)$.

To reduce to four-dimensional Einstein frame, we use the ansatz
\eqn\ansatz{ds_{10}^2 = e^{-6\sigma} ds_{4}^2 + e^{2 \sigma}
g^{0}_{mn}dy^m dy^n} where $g^0_{mn}$ is a fixed fiducial metric
normalized to have volume $4\alpha'^3$.  Although this differs
from the usual convention \eqn\usualansatz{ds_{10}^2 =
e^{-6(\sigma-\sigma_0)} ds_{4}^2 + e^{2 \sigma} g^{0}_{mn}dy^m
dy^n} by a constant rescaling, \ansatz\ is nevertheless
appropriate for a discussion of moduli stabilization, as we do not
know what the vev $\sigma_{0}$ will be until we stabilize
$\sigma$.  For a similar reason, we go between ten-dimensional
string and Einstein frame with the unconventional scaling
$g^S_{MN} = g^E_{MN} e^{{\phi} \over 2}$, while one usually sees
$g^S_{MN} = g^E_{MN} e^{{\phi-\phi_{0}} \over 2}$\Polchinski. The
resulting Minkowski metric differs from the conventional
$diag(-1,1,1,1)$ by a constant scaling depending on the vevs of
the dilaton and volume modulus. To relate dimensionful quantities
here to those directly measured from experiments, one must perform
an inverse rescaling. Finally, note that the gamma matrices built
from the metric scale with $e^\sigma$.

Let us decompose the ten-dimensional Majorana-Weyl gaugino
$\chi_{10}$ as \eqn\spinor{\chi_{10} = \chi_{6}^* \otimes \chi_{4}
+ \chi_{6}\otimes \chi_{4}^* } where $\chi_{6}$ and $\chi_4$ are
six and four-dimensional Weyl spinors with positive chirality and
$\chi_6$ is the zero mode of the internal Dirac operator for the
gaugino, with the normalization \eqn\normis { {\chi_{6}}^{\dag}
\chi_{6} = 1 .} We will choose to express the action in terms of a
rescaled four-dimensional gaugino $\lambda$
 \eqn\chiis{\lambda\equiv \chi_{4} e^{-{9\over{2}}\sigma
+ {\phi\over{4}}}} which will give the standard kinetic term after
dimensional reduction.

\bigskip {\noindent {\it{Coupling Constants}}}

The four-dimensional gauge coupling is \eqn\gymviav{ g_{YM}^2
\equiv  e^{\varphi}. } where the four-dimensional dilaton
$\varphi$ is related to the ten-dimensional dilaton and volume
modulus via \eqn\dilaton{\varphi = {\phi\over{2}}-6\sigma} Another
important scalar field of the four-dimensional theory is the
volume scalar\foot{For the moment we assume that the Calabi-Yau
has only one volume modulus.  We will present the more general
case in \S5.2.} $\rho$, \eqn\volscal{\rho =
{\phi\over{2}}+2\sigma} The fields $\varphi$,$\rho$ are related to
the scalar components of two ${\cal{N}}=1$ chiral superfields $S$,
$T$: \eqn\sandt{\eqalign{ S &= e^{-\varphi} + ia \cr T &= e^{\rho}
+ ib \cr }} where $a$ and $b$ are the axions which arise from the
spacetime and internal components of $B_{AB}$, respectively.  In
particular, \eqn\axion{ {\left(*da\right)}_{\mu\nu\rho} =
e^{-2\varphi}H_{\mu\nu\rho}} with an analogous relation for $b$.

The holomorphic Wilsonian gauge coupling functions $f_{i}^{W}$
(where $i=1,2$ runs over the two $E_8$ gauge groups) can be
expressed in terms of $S$ and $T$ by \eqn\wilson{ f_{i}^{W}= S +
{\beta}_{i} T + {\cal{O}}(e^{-S}) + {\cal{O}}(e^{-T})} where the
coefficient $\beta_i$ represents the one-loop correction to the
gauge coupling function, and the last two terms represent
nonperturbative corrections. Higher loop corrections vanish by
standard holomorphy arguments, since the dilaton and radion are
partnered in chiral multiplets with axions. The physical effective
coupling differs from the Wilsonian coupling by wave-function
renormalization and integration over the low momentum modes.

\bigskip {\noindent {\it{Four-dimensional Action}}}

Combining the relations given above, we reach the
dimensionally-reduced action \foot{ The unusual gravitational
coupling $\kappa_4^2 = {{\alpha'}\over{4}}$ is an artifact of our
ansatz \ansatz.  The physical gravitational coupling differs from
this by the constant rescaling mentioned previously.}
\eqn\stot{ S_{4d} = S_{gravity} + S_{gauge} + S_{CY} } \eqn\sgrav{
S_{gravity} = {2\over{\alpha'}}\int d^{4}x
\sqrt{-g_{4}}\left({{\cal{R}}_{4}-{{1\over{2}}{\partial_{\mu}\varphi
\partial^{\mu}\varphi}}}-{{3\over{2}}{\partial_{\mu}\rho
\partial^{\mu}\rho}}
\right)}

\eqn\sgauge{\eqalign{ S_{gauge} = & \int d^{4}x \sqrt{-g_{4}}
\left(-{1\over{2g_{YM}^2}}{\rm{tr}}(F_{\mu\nu}F^{\mu\nu})
-{2\over{g_{YM}^2}}
{\rm{tr}}(\overline{\lambda}\Gamma^{\mu}D_{\mu}\lambda)\right)}}


%
\eqn\scy{ S_{CY} =  { -{1\over{24\alpha'^4}}\int d^{4}x
\sqrt{-g_{4}}e^{\varphi-3\rho}\int_{X}d^6y \sqrt{-g^{0}} \left(
H_{lmn}- {\alpha'\over{16}}e^{12\sigma} T_{lmn}\right)^{2}}} where
we have defined \eqn\tis{T_{lmn} = \tr \left(({\chi_{6}}^{\dag}
{\bar{\lambda}_D}^* + {\chi_6}^T
\bar{\lambda}_D)\Gamma^{0}_{lmn}({\chi_{6}}^*\lambda_D +
{\chi_{6}} {\lambda_D}^* ) \right)} and $\lambda_D$ is the Dirac
spinor corresponding to $\lambda$.  The perfect-square interaction
term \scy\ couples the background flux to the gauginos and
therefore gives rise, as we will see in detail, to a potential for
the dilaton.

\bigbreak
\bigskip {\noindent {\it{Gaugino Condensation}}}

Recall that in a pure ${\cal{N}}=1$ supersymmetric Yang-Mills
theory in four dimensions with gauge group $H$, the gaugino
condensate which develops at low energies is given by
\refs{\shifv,\shifvtwo, \seiberg, \pouliot}:
\eqn\wilsoncon{{\Bigl< \tr\left({1 \over2 }\bar{\lambda}_D ( 1 -
\gamma_5) \lambda_D\right) \Bigr>} = {\langle \tr
(\lambda_{\alpha} \lambda^{\alpha }) \rangle} = {{16\pi^2}M^3
{\rm{exp}}\left(-{{8 \pi^2 f^{W}} \over{C_H }}\right)}.} Here $M$
is the ultraviolet cutoff for the gauge theory, $f^W$ is given by
\wilson, and $C_{H}$ denotes the dual Coxeter number of $H$. We
are interested in studying a gaugino condensate in some subgroup
$H$ of the hidden sector $E_8$ gauge group which arises in
compactification of the $E_8 \times E_8$ heterotic string on a
Calabi-Yau manifold. The appropriate ultraviolet cutoff $M$ for a
string compactification is the mass scale of Kaluza-Klein
excitations, \eqn\kkis{ M^3 = c
\left({e^{-12\sigma}\over{2{\alpha'}^{3/2}}}\right)} where $c$ is
a constant of order one.  Combining \wilsoncon\ and \kkis, we find
that the gaugino condensate in $H \subset E_8$ satisfies
\eqn\econdense{ {\langle \tr (\lambda \lambda )\rangle} =
{8\pi^2}c \left({e^{-12\sigma}\over{{\alpha'}^{3/2}}}\right)
{\rm{exp}}\left(-{{8 \pi^2 f^{W}} \over{C_H }}\right)}

\subsec{Superpotential from Flux and a Gaugino Condensate}

For a variety of reasons it will prove useful to work with a superpotential
and K\"{a}hler potential from which one can reproduce the interaction \scy.

One can derive the kinetic terms in \sgrav\ using the K\"{a}hler
potential \eqn\kahler{ {\cal{K}} = - \log(S+\bar S)-3 \log(T+\bar
T)-\log(-{i \over {4\alpha'^3}}\int \Omega\wedge\bar\Omega ).}

The superpotential for this system takes the form \eqn\wis{ W =
W_{flux} + W_{condensate} } where the first term is induced by the
background flux and the second term is a nonperturbative
contribution arising from the gaugino condensate.

The flux-induced superpotential can be written as an integral over
the Calabi-Yau space \refs{\GVW,\calibr,\Bhet,\BG} \eqn\wfis{
W_{flux} = {2 \sqrt{2} \over {\alpha'^4}} \int H\wedge\Omega} This
superpotential leads to the following term in the scalar potential
\eqn\vw{V_{flux} ={1\over{24\alpha'^4}}
e^{\varphi-3\rho}\int_{X}d^6y\sqrt{-g^0}H_{lmn}H^{lmn}} which is
precisely the first term in \scy.  As we will explain in \S3, the
number of quanta of H-flux is roughly given by \eqn\h{ h = {1
\over { 4 {\pi}^2 \alpha'^4 }} \int H \wedge \Omega } so that we
may define a mass parameter $\mu$, \eqn\muis{\mu^3 =  {{4 \sqrt{2}
c {\pi}^2  } \over {\alpha'^{3/2}}} }in terms of which \eqn\muw{
W_{flux} = \left({2\mu^3\over{c}}\right)h}

The nonperturbative contribution is conveniently expressed in
terms of the Wilsonian coupling \lk\ \eqn\wcondensate{
W_{condensate} = -C_H {\mu}^3 {\rm{exp}}\left(-{{8 \pi^2 f^{W}}
\over{C_H }}\right)} where the normalization was obtained by
comparing to \scy.  Putting these two pieces together, the total
superpotential is \eqn\wtotal{{W} = \left({2\mu^3\over{c}}\right)h
- C_H \mu^3 \exp \left(-{{8 \pi^2 f^{W}} \over{C_H }}\right) .}

\subsec{Conditions for a Stabilized Dilaton} A potential for the
dilaton arises from the perfect-square interaction term \scy,
which couples the background flux to the gauginos.  To analyze
this expression we first observe that the gaugino bilinear
appearing in \scy\ is proportional to the covariantly constant
holomorphic three-form.  This follows from the fact that
$\chi_{6}$ is a gaugino zero mode on the Calabi-Yau manifold
\drsw: \eqn\spingamma{\tr \left(({\chi_{6}}^{\dag} {\bar{\chi}}^*
+ {\chi_6}^T {\bar{\chi}})\Gamma^{0}_{lmn}(\chi_{6}^*\chi +
{\chi_{6}} {\chi}^* ) \right) = 2 \langle \tr(\lambda \lambda
)\rangle \Omega_{lmn} + c.c. } Here $\Omega$ is the holomorphic
$(3,0)$ form on the Calabi-Yau, with the normalization $ {1 \over
3 !} \Omega_{ijk} {\bar{\Omega}}^{ijk} = 1 $.

Minimizing the perfect square \scy\ forces $\langle \lambda \lambda \rangle
\Omega$ + $\langle \lambda \lambda \rangle^{*} \bar{\Omega}$ to align itself
along the same direction in $H^3(M, R)$ as the three-form flux
$H$. This uniquely fixes the complex structure moduli and the
four-dimensional gaugino condensate. Because the gaugino
condensate depends on the four-dimensional dilaton, it follows
that the interaction \scy\ generates a potential for the dilaton.

However, the minimum of this potential is generically at infinite
coupling. In the absence of Chern-Simons contributions, the
three-form $H$ obeys the quantization condition \eqn\bquanta{{1
\over {2{\pi}^2\alpha'}} \int_{Q}dB = n} for any $Q$ in
$H_{3}(X,\Z)$. The second term inside the perfect square of \scy,
on the other hand, integrates over three-cycles to
\eqn\gauginoint{\eqalign{ & {  \int_{Q}
{\alpha'e^{12\sigma}\over{8}} \Bigl( \langle \tr(\lambda \lambda)
\rangle \Omega_{ijk} + c.c. \Bigr) } \cr
 = ~& {{c{\pi}^2\over{\alpha'^{1/2} }} \exp
\left(-{{8 \pi^2} \over{C_{H} g_{YM}^2 }} \right) \left( e^{ - i
\theta} \int_Q \Omega + e^{ + i \theta}  \int_Q \bar{\Omega}
\right)}  \cr  ~\simeq~ & {{c} {\pi}^2 \alpha'
 \exp \left(-{{8 \pi^2} \over{C_{H} g_{YM}^2 }} \right)}}}
These two terms cancel only if
\eqn\vanish{{c \over 2} \exp \left(-{{8 \pi^2} \over{C_{H}
g_{YM}^2 }} \right) \simeq n. }  This has no solution because the
left hand side is almost always\foot{We are assuming that the
constant $c$ in \kkis\ is of order one.  If $c$ takes a larger
value in a particular model then integral flux might possibly
stabilize the dilaton, albeit at relatively strong coupling.  We
will not investigate this possibility here.} less than one. This
means that instead of stabilizing the four-dimensional dilaton at
a finite value, turning on an integral flux $dB$ actually drives
the system to infinitely strong coupling. Our proposal is to use
{\it{fractional}} fluxes to overcome this problem and stabilize
$g_{YM}$ at finite coupling. We therefore turn to an investigation
of the conditions under which fractional flux can arise in the
heterotic string.


\newsec{Fractional Flux Induced by Gauge Fields}

In \S3.1 we review the quantization condition for three-form flux
and explain its relation to the Chern-Simons invariant.  In \S3.2
we briefly discuss the class of three-manifolds used in our models
and construct a simple example. In \S3.3 we provide expressions
for the Chern-Simons invariants of these manifolds.  In \S3.4 we
discuss the conditions under which fractional Chern-Simons flux
leads to a worldsheet anomaly, and we explain how this can be
avoided in our setup.

\subsec{ Quantization Conditions for Three-Form Flux}

Consider a compactification of the $E_8 \times E_8$ heterotic
string on a Calabi-Yau manifold $X$. The two-form $B_{\mu\nu}$ is
required to satisfy
\eqn\bquantb{{1 \over {2 {\pi}^2 \alpha'}} \int_{Q}dB =  n}
for any three-cycle $Q$ in $H_{3}(X,\Z)$ in order for the action
of worldsheet instantons to be single-valued \rohmw. However, the
gauge invariant field strength is
\eqn\his{ H = dB - {\alpha'\over{4}}\Omega_{3}(A)
+{\alpha'\over{4}}\Omega_{3}(\omega) .}
This does not need to obey the same quantization law, due to the
presence of the Chern-Simons term. To see this let us assume for
simplicity that the background $B$-field is trivial, and that the
contribution of the spin connection $\omega$ can be ignored. Then
only the remaining factor of the gauge connection contributes. So
instead of \bquantb\ we find the quantization rule
\eqn\hquant{{1\over{2\pi^2\alpha'}}\int_{Q} H = - CS(A,Q) }
where we introduced a standard notation
\eqn\csis{\eqalign{ CS(A,Q) & = {1\over{8\pi^2}}
\int_{Q}\Omega_{3}(A)  \cr & ={1\over{8\pi^2}} \int_{Q}
\tr\left(A\wedge dA+{2\over{3}}A\wedge A\wedge A\right)}}
for the Chern-Simons invariant associated with a three-manifold
$Q$ and a connection one-form $A$.

The invariant $CS(A,Q)$ plays an important role in the theory of
three-manifolds. In particular, if $V'$ is a gauge bundle over $Q$
and if $A$ is a flat gauge connection on $V'$, then $CS(A,Q)$ is a
topological invariant, in the sense that $CS(A,Q)$ takes a fixed
value on each component of the moduli space of flat connections on
$Q$. Moreover, it is well known that $CS(A,Q)$ is well defined
only modulo integers and can take fractional values. If we further
assume that the bundle $V'$ pulls back to a gauge bundle $V$ over
the Calabi-Yau manifold $X$, then we obtain the desired situation
where the three-form flux takes fractional values. In the
following sections we will use this as a mechanism to produce
small quanta of the $H$-flux, which can then be used to stabilize
the various moduli.

\subsec{ Three-cycles with Fractional Flux }

Certain classes of three-cycles in Calabi-Yau manifolds admit
connections with fractional Chern-Simons invariant. We now turn to
a discussion of the properties of such three-cycles.

Since only holomorphic and antiholomorphic components of the
three-form flux contribute to the superpotential \wfis, the only
fractional fluxes we need to consider are those of Hodge type
$(3,0)+(0,3)$.  These can be viewed as fluxes through special
Lagrangian cycles $Q$. Typically these are compact three-manifolds
with non-negative curvature which support gauge fields suitable
for our purposes. According to McLean \mclean, the deformations of
a special Lagrangian submanifold $Q$ can be identified with the
harmonic one-forms on $Q$. Specifically, the deformation space has
real dimension $b_1 (Q)$. Therefore, rigid special Lagrangian
three-cycles are precisely rational homology three-spheres, i.e.
three-manifolds with $b_1 (Q) = 0$.  We shall henceforth restrict
our attention to rigid special Lagrangian three-cycles.  The local
Calabi-Yau geometry near such cycles is always of the form,
$$
T^* Q
$$
For example, we can choose $Q$ to be the base of the special
Lagrangian torus fibration \SYZ,
\eqn\syzfibr{f \colon X \to Q}

Indeed, following Strominger, Yau, and Zaslow \SYZ, consider a BPS
state in the effective four-dimensional theory represented by $N$
D6-branes wrapped over the entire mirror manifold $\tilde X$.
These D6-branes are rigid and, because the fundamental group of
$\tilde X$ is finite, there is only a discrete set of Wilson
lines. In fact, the latter account for the degeneracy of D-brane
bound states \GV. Namely, the number of bound states of $N$
D-branes is given by the number of $N$-dimensional irreducible
representations of $\pi_1 (\tilde X)$. Under mirror symmetry
(realized as T-duality on $T^3$ fibers) these D6-branes become
D3-branes wrapped around the base $Q$. In order for the D3-branes
to have no continuous moduli the base manifold $Q$ must be a
rational homology three-sphere. Also, by looking at the degeneracy
of D-brane bound states for different values of $N$, we conclude
that $\pi_1 (Q)$ and $\pi_1 (\tilde X)$ should be related.
Notice that since both $X$ and its mirror $\tilde X$
are fibered over the same base $Q$, the above arguments
imply that their homotopy groups should be related as well.
In particular, in a large class of examples one finds
that the abelian parts of $\pi_1 (X)$ and $\pi_1 (\tilde X)$
are isomorphic,  cf. \Aspinwall.

Let us study a simple example that will be relevant in the
following. Consider a quintic hypersurface in ${\bf CP}^4$,
\eqn\quintic{z_1^5 + z_2^5 + z_3^5 + z_4^5 + z_5^5 + {\rm
~deformations~} =0 }
This hypersurface represents a Calabi-Yau variety $X_0$ with
$h^{1,1}=1$, $h^{2,1} = 101$. Unfortunately, $\pi_1 (X_0)$ is
trivial, so $X_0$ does not admit a fractional flux induced by
non-trivial gauge fields. Moreover, since the number of
generations in a heterotic compactification on a Calabi-Yau
threefold $X$ is related, in the case of the standard embedding,
to the Euler number of $X$ \CY, in the present case with the
standard embedding we find an
unrealistically large number, $N = {1 \over 2} \vert \chi (X_0)
\vert =100$. A model with a more realistic spectrum that does not
suffer from these problems can be obtained by considering a
quotient of $X_0$,
$$
X = X_0 / \Gamma
$$
by a discrete symmetry group $\Gamma = \Z_5 \times \Z_5$,
generated by two elements
\eqn\quinticg{\eqalign{ g_{1} & \colon (z_1, z_2, z_3, z_4, z_5) \to
(z_5, z_1, z_2, z_3, z_4) \cr g_{2} & \colon (z_1, z_2, z_3, z_4, z_5)
\to (\zeta z_1, \zeta^2 z_2, \zeta^3 z_3, \zeta^4 z_4, z_5) }}
where $\zeta = \exp (2\pi i/5)$. Since $\Gamma$ acts freely on
$X_0$, we have $\chi (X) = \chi (X_0) /25 = 8$ and $\pi_1 (X) =
\Z_5 \times \Z_5$. Therefore, compactification of the heterotic
string on the resulting manifold $X$ with the standard
embedding provides a model with only
four generations, and there is a possibility to turn on
non-trivial Wilson lines on $X$. Also, it is easy to see that the
base, $Q$, of the special Lagrangian torus fibration in this case
is a rational homology three-sphere with non-trivial fundamental
group.

For the quintic hypersurface \quintic, the base $Q_0$ of the
special Lagrangian torus fibration can be represented by the image
of the moment map, $z_i \to \vert z_i \vert^2$. The topology of
$Q_0$ can easily be understood in the large complex structure
limit, where it is close to the boundary of the toric polytope.
Hence, $Q_0 \cong {\bf S}^3$. Now let us consider the action of
the discrete group $\Gamma$. {}From \quinticg\ it follows that the
generator $g_{2}$ acts trivially on $Q_0$, whereas $g_{1}$ acts freely.
Therefore, we find that the base of the special Lagrangian torus
fibration $X \to Q$ is a Lens space,
\eqn\qzfive{ Q = {\bf S}^3 / \Z_5 }
In particular, we have $\pi_1 (Q) = \Z_5$ and, as we will show
below, there are many choices for the gauge bundle $V'$ and for
the gauge connection $A$ over this three-manifold, such that
$CS(A,Q)$ has fractional values. If $V'$ is such a bundle, we can
define its pullback $V = f^{-1} V'$ under the projection map
\syzfibr. The resulting gauge bundle $V$ over $X$ has the desired
properties and, according to the quantization rule \hquant, the
three-form flux in heterotic string theory on this background can
take fractional values.

This construction can easily be generalized to an arbitrary
special Lagrangian three-cycle $Q$ which is rigid inside $X$. As
was explained above, the condition of rigidity implies that $Q$ is
a rational homology three-sphere. Examples of rational homology
three-spheres that can occur as special Lagrangian cycles in
Calabi-Yau threefolds include Lens spaces, Brieskorn homology
three-spheres, and, more generally, Seifert fibered
three-manifolds. Recall that the Seifert three-manifold, $\Sigma
(a_1, \ldots, a_n)$, is a circle fibration over a two-sphere, with
$n$ multiple fibers. This includes Brieskorn spheres and Lens
spaces as a special case, $n=3$. For instance, the Lens space
$L(p,1) = {\bf S}^3 / \Z_p$ is a Seifert three-manifold with
$(a_1,a_2,a_3) = (p,2,2)$. Many of these three-manifolds support
non-trivial gauge connections with fractional Chern-Simons
functional \refs{\nishi,\auckly}.

\subsec{Formulas for the Chern-Simons Invariant}

In order to determine the set of values of $CS(A,Q)$ for a given
three-manifold $Q$, one has to study the space of representations
of the fundamental group, $\pi_1 (Q)$, into the gauge group.
A familiar example of a reducible\foot{A connection $A$ is called
reducible if its isotropy subgroup, that is a maximal subgroup
that commutes with all the holonomies of $A$, is a continuous
group. Otherwise, $A$ is called irreducible. For example, an
$SU(2)$ gauge connection is reducible if its isotropy subgroup is
$U(1)$. Notice that reducible gauge connections may have non-zero
Chern-Simons invariant, see {\it e.g.} \Rozanskyseif.
} gauge connection on a manifold with $\pi_1 = \Z_p$
corresponds to a discrete Wilson line of the form
\eqn\uis{ U = {\rm{diag}}(e^{2\pi i k_1/p},\ldots,e^{2\pi i
k_8/p})}
variations of which are often used to break the GUT gauge group to
a smaller subgroup, such as the Standard Model gauge group \Wsb.
The Chern-Simons invariant of such a connection is \anomaly~(see
also \conrad)
\eqn\inv{ CS(A,Q) = \sum_{i} { k_i^2\over{2p}} \quad\quad
\rm{mod}~~\Z }
where the sum is over all eight complex worldsheet fermions. For
appropriate choices of $p$ and of the $k_{i}$ the result is a
fractional Chern-Simons invariant.\foot{In \S3.4 we review the
existence and cancellation of a potential worldsheet global
anomaly in such backgrounds.}

This has the surprising consequence mentioned in the introduction:
in many cases the Wilson lines which are used to break the GUT
gauge group to the Standard Model introduce a fractional
Chern-Simons invariant, and hence a fractional flux.

We now turn to the more general question of the fractional
Chern-Simons invariants of Seifert three-manifolds; this choice
covers a fairly large class of models relevant to the physical
problem at hand.  Without loss of generality, we can take the
gauge group to be $SU(2)$ (which can be realized as a subgroup in
one of the two $E_8$'s).  Let $Q = \Sigma (a_1, \ldots, a_n)$ be a
Seifert three-manifold.  In this case, the irreducible
representations,
$$
\rho : \quad \pi_1 (Q) \to SU(2)
$$
are characterized by what are called ``rotation numbers'', $(\pm
m_1 , \ldots, \pm m_n)$, where each $m_i$ is defined modulo $a_i$,
$$
m_i \sim m_i + a_i
$$

Furthermore, there exists at most one component of the
representation variety realizing a given set of rotation numbers
$(m_1, \ldots, m_n)$. If $A$ is the corresponding connection
one-form, the value of the Chern-Simons functional, $CS(A,Q)$, is
given by the simple formula
\eqn\seifertcs{ CS(A,Q) = - \sum_{i=1}^3 {1 \over a_i} \left(m_i +
\lambda \right)^2 }
where
\eqn\lambdacases{\lambda = 0 ~,~ {1 \over 2}}
In particular, if $Q = {\bf S}^3/\Z_p$ is a Lens space, from the
general formula \seifertcs\ we find
\eqn\cslens{CS(A, Q) =
- {1 \over p} \left( m_1 + \lambda \right)^2
- {\lambda^2 \over 2} \quad \quad {\rm mod~~} \Z }
where for simplicity we set $m_2 = m_3=0$. This expression gives
two sets of values of the Chern-Simons functional (listed in
\nishi) corresponding to $\lambda=0$ and $\lambda = 1/2$,
respectively. It is convenient to introduce a new integer
parameter
$$
m = 2m_1 + 2 \lambda \quad \quad {\rm mod~~} 2p
$$
and rewrite \cslens\ in the form
\eqn\lenscs{CS(A,Q) = - {m^2 \over 4p} - {\lambda^2 \over 2} \quad
\quad {\rm mod~~} \Z. }
In general, it follows from \seifertcs\ that $CS(A,Q)$ is a
rational number whose denominator can be as large as the order of
the fundamental group, $\pi_1 (Q)$.

\subsec{A Global Worldsheet Anomaly from Fractional Chern-Simons
Invariants}

For completeness, we now discuss a technical issue related to
modular invariance in a fractional flux background. Specifically,
we present a sufficient condition for cancellation of the
worldsheet anomaly induced by fractional Chern-Simons
flux.\foot{We are indebted to E. Witten for explaining to us
much of the content in this subsection.}

When the heterotic string propagates on a nontrivial geometry $M$
with nontrivial Wilson lines, there is a global worldsheet anomaly
in addition to the one-loop anomaly seen in the ten-dimensional
supergravity \anomaly.  This signals that the worldsheet instanton
path integral is not necessarily single-valued in such a
background.

To compute the anomaly, consider a one-parameter ($t$) family of
maps from a one-parameter family of worldsheets into the target
space, with the worldsheets at $t=0$ and $t=1$ identified by a
large diffeomorphism $h$ preserving the spin structure: $ \varphi:
(\Sigma \times [0, 1]_t)_h \rightarrow M$. The change of the
fermion determinant can be calculated using an index theorem
\anomaly, \eqn\effectiveaction{\ln Z(\phi^i, t=1) - \ln Z(\phi^i,
t=0) =- 2 \pi i \int_ { \varphi(\Sigma \times [0, 1])_h }\Omega_3
(A),} where \eqn\determinant{ Z(\phi^i, t; g_{ij}, B_{ij},
{A_i^A}_B)=
({\det}^+_T)({\det}^-_{V_1})({\det}^-_{V_2})({\det}^+R).} Here the
first three terms inside the logarithm are Dirac determinants for
the right- and left-moving fermions coupled to the pull-back of
the spin connection and gauge connection, and the fourth term
comes from the right-moving Rarita-Schwinger ghost. If we were
unable to find other sources to cancel the factor on the right
hand side, we would have to set the Chern-Simons invariant to an
integer to maintain the single-valuedness of the determinants.

Fortunately the Wess-Zumino term on the worldsheet can help us.
For the heterotic string on a Calabi-Yau with flat $B$ field and with
no Wilson lines, the worldsheet action looks like
\eqn\worldsheetaction{\eqalign{S = & {\int d^2x \biggl(
\bigl(g_{ij}(\phi)+B_{ij}(\phi)\bigr)\partial_+ \phi^i
\partial_- \phi^j + i g_{ij} \psi^i \bigl( \partial_- \psi^j +
\Gamma^i_{jk}
\partial_- \phi^k \psi^l \bigr)\biggr. }
\cr &  \biggl. + i G_{AB}(\phi) \lambda^A \Bigl(\partial_+
\lambda^B + {{A_i}^B}_C (\partial_+ \phi^i) \lambda^C \Bigr) + {1
\over 2} \widetilde{F}_{ijAB}\psi^i \psi^j \lambda^A \lambda^B
\biggr)}} where $\psi^i$ and $\lambda^A$ are the right- and
left-moving fermions, $\Gamma^i_{jk}$ is the Levi-Civita
connection of the target space, and $G_{AB}$ is the metric on the
gauge bundle. This action has manifest (0,2) supersymmetry.  The
question is, if we now turn on flat Wilson lines supporting
fractional Chern-Simons invariant, resulting in multi-valued
fermion determinants, can we find cancelling effects from the
bosonic worldsheet action? The answer is {\it{yes, provided there
is no torsion in}} $H^4(M,\Z)$.

To see this, consider the following exact sequence:
\eqn\exactsequence{
\cdots \mapright{} H^3(M, {\bf R}) \mapright{e}
H^3(M,U(1)) \mapright{d} H^4(M, \Z) \mapright{} H^4(M, {\bf R})
\mapright {} \cdots .}
The Chern-Simons invariant $\exp(i \int
\Omega_3(A))$ for a flat bundle takes values in $H^3(M,U(1))$ and
is mapped into the torsion part of $H^4(M, \Z)$. If $H^4(M, \Z)$
is torsion-free, the Chern-Simons invariant lives in the kernel of
$d$ and therefore $\Omega_3(A)$ lives in $H^3(M, {\bf R})$. So there
exists, locally, a two-form $\widetilde{B}$:
$$ d \widetilde{B} = \Omega_3(A) .$$
It is crucial that $\widetilde{B}$ is not globally defined when $\int \Omega_3(A)$
is fractional. The change in phase from the coupling of
$\widetilde{B}$ to the worldsheets cancels the change in the
fermion determinants in equation \effectiveaction\worldsheet. On
the other hand, if $H^4(M, \Z)$ has a torsion piece,
$\widetilde{B}$ does not exist for bundles supporting fractional
Chern-Simons invariant and we cannot cancel the global worldsheet
anomaly. The only consistent Wilson lines are then those that give
integer Chern-Simons fluxes.

The reader will have noticed that if we modify the Wess-Zumino
term into $$\int_{\Sigma} B + \widetilde{B},$$ we no longer have
(0,2) worldsheet supersymmetry. We can preserve (0,1)
supersymmetry by modifying the connection to
\eqn\modifyconnetion{\widetilde{\Gamma}^i_{jk} ={\Gamma}^i_{jk} +
g^{il} (d \tilde{B})_{jkl} =\Gamma^i_{jk} + g^{il} \Omega(A)_{jkl}.}
However, the complex structure ${J^i}_j$ is no longer covariantly
constant.  Thus, just as we expected, turning on a flat bundle
with Chern-Simons gauge flux generates a spacetime superpotential
$W = \int \Omega_3(A) \wedge \Omega$ and breaks ${\cal{N}}=1$
spacetime supersymmetry and (0,2
) worldsheet supersymmetry. It is
obvious from the supergravity effective action that with the
addition of a gaugino condensate, spacetime supersymmetry can be
restored.  However, we do not expect a useful worldsheet
description after including such spacetime
effects.\foot{Alternatively, to preserve (0,2) worldsheet
supersymmetry, one could modify ${J^i}_j$ so that
$\widetilde{\nabla}_i {J^j}_k = {J^j}_{k,i} +
\widetilde{\Gamma}^j_{il}{J^l}_k -
\widetilde{\Gamma}^l_{ik}{J^j}_l =0 $ with respect to the modified
connection. This typically cannot be achieved by a local
modification (i.e. a continuous deformation) and requires starting
with a non-K\"ahler manifold. This is closely related to \torsion\
and to more recent literature on non-K\"ahler compactifications.
The difference is that here we would consider non-K\"ahlerity due
to $\Omega_3(A)$ instead of the more conventional non-flat $d B$.}

We have seen, then, that a sufficient condition for cancellation
of the worldsheet anomaly in the presence of fractional flux is
absence of torsion in $H^4(M, \Z)$.  More specifically, it is
enough that no three-cycle $Q$ on which the Chern-Simons form
integrates to a fraction is a torsion cycle in $H_{3}(M,\Z)$.  We
will henceforth assume that this condition is satisfied.


\newsec{Dilaton Stabilization}

We will now demonstrate that the combination of a gaugino
condensate and a fractional flux induced by the Chern-Simons
term of the $E_8 \times E_8$ gauge connection can lead to
stabilization of the dilaton at finite (and, with sufficient
tuning, weak) coupling.

We denote the two gauge groups $E_8^{(i)}$, $i=1,2$. Let us henceforth
adopt the convention that $E_8^{(1)}$ is the observable $E_8$ and
$E_8^{(2)}$ is the hidden sector.  We imagine that there is a suitable
visible-sector bundle which breaks $E_8^{(1)}$ to an attractive GUT group.
If a realistic model is desired, we may also require that the observable
$E_8^{(1)}$ has a gauge bundle with $\vert \int c_{3} \vert = 6$ to give
three generations of quarks and leptons.\foot{Examples of Calabi-Yau
models with three generations and nontrivial $\pi_1$ have appeared in
\threegen, and undoubtedly many more could be constructed in a systematic
search.} In the remaining visible-sector group we then turn on Wilson
lines which have fractional Chern-Simons invariant on some three-cycle.
The resulting fractional flux generates a superpotential via
\wfis.\foot{The fractional flux could instead come from hidden-sector
Wilson lines.  We focus on visible-sector Wilson lines for simplicity.}

For the purposes of this section we could take the hidden-sector
bundle to be trivial, so that $E_8^{(2)}$ is unbroken.  However,
it will prove useful in \S5\ to include a non-trivial gauge bundle
in each of the $E_8$s.  We therefore embed an $SU(2)$ bundle into
$E_8^{(2)}$, breaking $E_8 \to E_7$. There is no index theorem
protecting charged matter in $E_7$ (as it has only real
representations), so we can safely assume that the low-energy
$E_7$ gauge theory in the hidden sector has no light ${\bf 56}$s.
The gauge group then confines at low energies, providing a gaugino
condensate to balance the fractional flux, as in \S2.2.

The overall result is the superpotential \wtotal:
\eqn\Wnowis{{W\over{\mu^3}} = {2 h \over c} ~-~ 18 ~{\rm{exp}}
\left(- {8 \pi^2 S \over 18} \right) } where $h = {( 2 {\pi}^2
\alpha'^{5/2})^{-1}} \int{H\wedge\Omega}$ is the flux contribution
and the second term is the result of gaugino condensation (the
dual Coxeter number of $E_7$ is 18).\foot{This superpotential is
of the same form as the one appearing in, for instance, equation
(12) of \kklt.  There, the small constant term comes from the
$(0,3)$ part of the type IIB $G_3$ flux, while the exponential
arises from nonperturbative gauge dynamics as in our system.}

To look for a supersymmetric vacuum, we solve the equation $D_{S}
W = 0$, with the result \eqn\solis{h~=~\Bigl(9c+ 8 c{\pi}^2  Re(S)
 \Bigr) {\rm{exp}} \left( -{{8\pi^2 S} \over 18}\right)~.} Modest
values of the Chern-Simons invariant lead to a solution at weak
coupling.  For example, if $h$ is approximately ${1\over 10}$,
which is easily attainable using the constructions of \S3, then
\solis\ can be solved with $Re(S) \sim 1.6$, which corresponds to
$\alpha_{GUT} \sim {1\over 20}$.  To achieve instead the
often-quoted value $\alpha_{GUT} \sim {1\over 25}$ one needs $h$
of order ${1\over 40}$.  Of course the requirements are weaker
if we take the pure hidden sector gauge group to be $E_8$ instead
of $E_7$.

There are many variations of this mechanism which involve slightly
different choices of bundles.  It seems to us that the most
elegant models are those in which one set of Wilson lines breaks
the observable-sector GUT group to the Standard Model and also
provides the needed fractional Chern-Simons invariant.

We have already solved the dilaton equation $D_S W = 0$.  We can
likewise solve the equations for the complex structure moduli by
making $H$ of type $(3,0) + (0,3)$.  In this way the $H$-flux from
the Chern-Simons invariant generically stabilizes all complex
structure moduli. The K\"ahler moduli of the Calabi-Yau, however,
are not yet fixed. In particular, there is a flat direction for
the volume modulus $T$.\foot{If there are vector bundle moduli
then these are also unfixed.  However, in \S8 we explain why
bundle moduli could be absent in generic situations.}

In fact, this flat direction is a general property of ``no-scale''
models. From the form \kahler\ of the K\"ahler potential, combined
with the fact that $W$ is independent of the volume modulus $T$ at
this order, we see that the supergravity potential undergoes a
simplification \eqn\supgropt{V = e^{K} \left(g^{i\bar j}D_i W
\overline{ D_{j}} \overline{W} - 3 \vert W \vert^2 \right) \to
e^{K} \left(g^{a\bar b} D_{a}W \overline{ D_{b}  W}\right)} where
$i,j$ run over all fields, but $a,b$ run over all fields ${\it
except} ~T$. As a result, we are left with a flat direction, $T$.
Generically $D_T W \neq 0$, so supersymmetry is broken.
Nevertheless, the vacuum energy vanishes at this order of
approximation, since we have solved $D_a W = 0$ for all $a$. Loop
corrections will plausibly destabilize $T$, resulting in a runaway
problem for the overall volume.

We will suggest a solution to this problem, in the context of
Calabi-Yau compactification, in the next section.  However, we
should point out that investigation of supersymmetric non-K\"ahler
compactifications of string theory has recently been renewed
(see e.g. \refs{\nonkahler,\bbdp,\bbd,\Cardoso}).
In such compactifications
the overall volume modulus can be stabilized at tree level by balancing
fluxes against the non-K\"ahler nature of the geometry.
The combination of this tree-level $T$ stabilization with our
results on dilaton stabilization could plausibly yield
weakly-coupled models with all moduli stabilized.  This would
require a compactification manifold which admits moderately small
Chern-Simons invariants.

\newsec{Dilaton ${\it and}$ Volume Stabilization in Calabi-Yau Models}

In \S5.1 we show that it is possible, with appropriate choices of
bundles, to stabilize both the dilaton and the overall volume by
incorporating the one-loop correction to the gauge coupling. In
\S5.2 we extend this mechanism to stabilize all the K\"ahler
moduli of a threefold. In \S5.3 we investigate the strong-coupling
transition which occurs in these models.  We present a toy model
to illustrate the physical smoothness of this transition.  In
\S5.4 we discuss the conditions under which the resulting theory
is weakly coupled. In \S5.5 we summarize our assumptions
concerning the Calabi-Yau and the $E_{8}$ gauge bundles.

\subsec{One-loop Correction}

We first consider, for simplicity, the case of a Calabi-Yau
threefold which has $h^{1,1}=1$ and hence a single K\"ahler
modulus.  When one-loop corrections are incorporated, the
Wilsonian gauge kinetic functions have the form \wilson:
\eqn\gf{f^{W}_{(i)} =  S + \beta_i T ,} where $i = 1,2$ labels the
gauge groups $E_8^{(1)}$, $E_8^{(2)}$. In the case without
space-filling heterotic five-branes, it is a simple matter to
derive the linear terms in $T$ by dimensional reduction of the $B
\wedge X_8(F_{1},F_{2},R)$ term in the ten-dimensional $E_8 \times
E_8$ theory. The result is \eqn\betaoneis{\beta_1 = {1\over
8\pi^2} \int_{X} J \wedge \Bigl( c_{2}(V_1) - c_{2}(V_2) \Bigr)~,}
\eqn\betatwois{\beta_2 = {1\over 8\pi^2} \int_{X} J \wedge
\Bigl(c_{2}(V_2) - c_{2}(V_1)\Bigr)~.} Here $J$ is the generator
of $H^{1,1}(X,\Z)$. Notice that \eqn\betasum{\beta_1 + \beta_2 =
0} while in the case of the standard embedding \eqn\betadiff{
\beta_1 - \beta_2 = {1\over 4\pi^2} \int_{X} J \wedge c_{2}(TX)~.}
This fact that the difference of the gauge coupling functions is
given by a topological invariant (in the case of the standard
embedding) was observed in e.g. \BCOV. One can easily calculate
$\beta$ for a few simple examples. We present the calculation
below for $J \wedge c_{2}(TX)$; one can imagine partitioning this
into $c_{2}(V_{1,2})$ in various ways.
$$
\int_{[4 \parallel 5]} J \wedge c_2
= 10 \int_{[4 \parallel 5]} J \wedge J \wedge J = 50,
$$
$$
\int_{[5 \parallel 3\,\,3]} J \wedge c_2
= 6 \int_{[5 \parallel 3\,\,3]} J \wedge J \wedge J =54,
$$
$$
\int_{[6 \parallel 3\,\,2\,\,2]} J \wedge c_2
=5 \int_{[6 \parallel 3\,\,2\,\,2]} J \wedge J \wedge J = 60.
$$
{}From these examples it is plausible that $\beta$ can be reasonably
large, at least of order one.

We will choose the gauge bundle $V_2$ so
that $E_8^{(2)}$ is broken to a subgroup $H$ (say $E_7$)
without any light charged
matter.  The resulting four-dimensional theory therefore has a
sector which is pure ${\cal N}=1$ supersymmetric gauge theory with
gauge group $H$, which undergoes gaugino condensation at low
energies.  Let us furthermore choose the bundle $V_1$ so that
$E_8^{(1)}$ is broken to a low-energy group and matter content
which can contain the Standard Model.  Finally, we take $\beta_2 =
-\beta_1 = \beta >0$, so that $E_8^{(1)}$ is more strongly coupled
than $E_8^{(2)}$.\foot{Notice that we are putting more instantons
in the hidden sector than in the observable sector, which is a
somewhat unusual situation compared to the bulk of the literature.}

The complete superpotential is then \eqn\wfinal{ {W\over{\mu^3}} =
{{2h}\over{c}} ~-~ {C_H}{\rm{exp}}\left(-{{8 \pi^2\over{C_H
}}(S+\beta T)}\right).}  This superpotential depends nontrivially
on both of the chiral multiplets $S$ and $T$. The condition for a
supersymmetric vacuum is \eqn\vacuum{ W;_S = W;_T = 0 } where the
K\"{a}hler covariant derivatives are determined using \kahler.

A solution of \vacuum\ necessarily satisfies \eqn\solutiona{3 S =
\beta T, } \eqn\solutionb{h = \left( {{C_H c}\over{2}}+ 8c{\pi}^2
Re(S) \right) {\rm{exp}}\left(-{{32 \pi^2 S\over{C_H }}}\right)}
The resulting solution is a supersymmetric AdS vacuum in which
both the four-dimensional dilaton $\varphi$ and the
four-dimensional volume modulus $\rho$ have been stabilized.
We will defer our discussion of the physics in $E_8^{(1)}$ to
\S5.3.

\subsec{ Stabilization of Multiple K\"ahler Moduli }

On a threefold X with $h^{1,1} > 1$ K\"ahler moduli, the formulas
of the previous section can be generalized: \eqn\kgf{f^{W}_{(i)} =
S + \beta_i^{\alpha} T_{\alpha} ,} where $i = 1,2$ labels the
gauge groups $E_8^{(1)}$, $E_8^{(2)}$ and $\alpha = 1, \ldots,
h^{1,1}$ indexes the independent K\"ahler moduli.

We will need to define a few quantities related to the generators
$J^{\alpha}$ of $H^{1,1}(X,\Z)$: \eqn\kbetaoneis{\beta_1^{\alpha}
\equiv {1\over 8\pi^2} \int_{X} J^{\alpha} \wedge \Bigl(
c_{2}(V_1) - c_{2}(V_2) \Bigr)~,} \eqn\kbetatwois{\beta_2^{\alpha}
\equiv {1\over 8\pi^2} \int_{X} J^{\alpha} \wedge \Bigl(c_{2}(V_2)
- c_{2}(V_1)\Bigr)~.} \eqn\intersect{c_{\alpha\beta\gamma} \equiv
\int_{X} J^{\alpha}\wedge J^{\beta}\wedge J^{\gamma}} The
$c_{\alpha\beta\gamma}$ are the intersection numbers of $X$.

The K\"{a}hler potential \kahler\ now takes the form
\eqn\kkahler{{\cal{K}} = - \log(S+\bar S)-\log \left
(c_{\alpha\beta\gamma}{\cal{T}^{\alpha}}{\cal{T}^{\beta}}{\cal{T}^{\gamma}}
\right)-\log\left(-{i\over{4\alpha'^3}}\int \Omega\wedge\bar\Omega
\right)} with $2 {\cal{T}^{\alpha}} \equiv T^{\alpha} +
\bar{T}^{\alpha}$, while the complete superpotential, including
hidden-sector gaugino condensation, is \eqn\kwfinal{
{W\over{\mu^3}} = {{2h}\over{c}} ~-~ {C_H}{\rm{exp}}\left(-{{8
\pi^2\over{C_H }}(S+\beta^{\alpha}T_{\alpha})}\right).} This
superpotential depends nontrivially on the dilaton and on all the
K\"ahler moduli.

In order to find a supersymmetric solution we will assume that all
the $\beta^{\alpha}$ are nonzero.
Combining \kwfinal\ and \kkahler\ and imposing $W;_S =
W;_{T_{\alpha}} = 0$, we find \eqn\ksolutiona{ S
{\partial\over{\partial{T_{\delta}}}}\left
(c_{\alpha\beta\gamma}{\cal{T}^{\alpha}}{\cal{T}^{\beta}}{\cal{T}^{\gamma}}
\right) = \beta_{2}^{\delta} \left
(c_{\alpha\beta\gamma}{\cal{T}^{\alpha}}{\cal{T}^{\beta}}{\cal{T}^{\gamma}}
\right)} \eqn\ksolutionb{h = \left( {{C_H c}\over{2}}+ 8c{\pi}^2
Re(S) \right) {\rm{exp}}\left(-{{32 \pi^2 S\over{C_H }}}\right)}
where the second relation is identical to \solutionb.

The result is a supersymmetric AdS vacuum without moduli. To
recapitulate, we have now seen that the combination of fractional
flux with a gaugino condensate can stabilize the complex structure
moduli, the K\"ahler moduli, and the dilaton.

\subsec{ A Strong Coupling Problem }

 We have just seen that the potential for the dilaton
and K\"ahler moduli has a supersymmetric AdS minimum whose
location is given, in the case of one K\"ahler modulus, by
\solutiona,\solutionb.  However, there is an evident problem with
this minimum.  Suppose that some subgroup of $E_8^{(1)}$ remains
unbroken at low energies. The naive $E_8^{(1)}$ gauge coupling
function, $f_{1} = S - \beta T$, appears to be {\it{negative}},
$f_1 = - 2S$.

Moreover, one might think that before becoming negative, $f_1$
must pass through zero, at which point one encounters a
singularity where the gauge coupling diverges.

It is clear a priori that such a problem cannot exist in the full
theory.  Moduli (and parameter) spaces of four-dimensional
supersymmetric theories are complex and hence can only have
singularities at complex codimension one.  It follows that one can
always continue around any point of naively singular gauge
coupling, obtaining a unitary theory with positive $g^2$ on the
``other side''.  Numerous examples of such phenomena have been
explored in various four-dimensional supersymmetric gauge theories
over the past several years, most recently in interpreting the
$G_2$ flop in \AMV.

In fact, what we are encountering here is (at least in those cases
which are most easily understood) a close relative of the
well-studied strong coupling transitions in six-dimensional string
vacua with (0,1) supersymmetry \sixd.  The observable sector gauge
coupling diverges precisely when the ratio $S/T$ reaches a fixed
value; this is in fact a point in moduli space where an effective
${\it six}$-dimensional coupling is becoming strong.  As explained
in \sixd, in dual type II or F-theory descriptions, this
phenomenon can be modeled locally in terms of a geometric
transition which affects the D-branes or local geometry
responsible for $E_8^{(1)}$. On the other side of the geometric
transition, the $E_8^{(1)}$ physics remains sensible, and there is
a new effective description of the low energy gauge theory.

In the remainder of this subsection we investigate this strong
coupling singularity.  The resolution is necessarily
model-dependent, so we simply review some dual descriptions which
shed light on the phenomenon, and give an explicit example where
the physics on the ``other side'' of the transition is fully
understood. Of course in as much as one wishes to embed the standard
model in $E_8^{(1)}$, it would be crucial to have a good dual
description of this new phase.  For readers who find this too
daunting a challenge, we can only suggest that the special case
$\beta_{1,2}=0$ neatly sidesteps the issue, leaving a no-scale
model with an unfixed volume modulus. However we emphasize that
more generally, the only assumption we really need to make is that
the physics of the transition does not introduce new terms in the
superpotential. For models where $E_8^{(1)}$ is broken to a
low-energy field theory that does not dynamically generate a
superpotential, this is quite plausible.

\bigskip\noindent {\it{ Dual Descriptions of the Strong Coupling
Singularity}}

\medskip The appearance of strong gauge coupling in
heterotic models with nonzero $\beta$ is well known.  The problem
is easily seen in compactifications of heterotic M-theory to four
dimensions, where it manifests as a linear shrinking \Wcy\ of the
Calabi-Yau volume as a function of location on the M-theory
interval.  For some critical size of the interval, the Calabi-Yau
has zero volume at one boundary, rendering the supergravity
approximation invalid.

A closely related problem arises in compactifications of the $E_8
\times E_8$ heterotic string on $K3 \times T^{2}$.  The gauge
bundle in such a model is specified in part by a choice of
instanton numbers $(12-n,12+n)$ in the two $E_8$s.  If $n$ is
positive then the first $E_8$ is more strongly coupled than the
second; this is analogous to positive $\beta$ in our models.  At a
finite value of the heterotic dilaton the first $E_8$ has infinite
gauge coupling.

This configuration is dual to compactification of type IIA string
theory on a Calabi-Yau threefold which is an elliptic fibration
over the Hirzebruch surface $F_{n}$. Recall that $F_{n}$ has a
single curve of self-intersection $-n$.  The volume of this curve
is dual to the heterotic dilaton in such a way that shrinking the
$(-n)$ curve to zero volume coincides with infinite gauge coupling
in the first $E_8$.  This suggests that one could use the type II
geometry to understand the nature of the strong coupling
singularity.  While this approach is rather complicated
for general $n$ (see e.g. \sixd\ for work in this direction),
we will see
that the case $n=1$ is relatively straightforward.

It is important to remember that type II strings on such a
Calabi-Yau threefold yield ${\cal{N}}=2$ supersymmetry in four
dimensions, twice as much as the models we have considered in this
paper.  This greatly facilitates analysis of the singularity, in
particular because the geometry can be described via a
prepotential.  A direct study of the ${\cal{N}}=1$ system would be
more challenging, but we expect the generic features, including
the positive gauge coupling function, to be similar in the two
cases.  One would simply have to study the geometry of a dual
F-theory compactification on a Calabi-Yau fourfold, instead of
type II strings on a Calabi-Yau threefold.

\medbreak\bigskip\noindent {\it{  A Simple Flop Model of the
Strong Coupling Singularity}}
\medskip

We will now construct a simple model in which, in a sense which we
will make precise, the gauge kinetic term $f_{1}$ undergoes a
flop.

Recall that in the flop of a curve, the volume of the curve
vanishes on a wall of the K\"ahler cone.\foot{In the full physical
theory the volume is complexified, and one can go ``around'' the
wall of the K\"ahler cone by turning on a nonzero $\theta$ angle
\refs{\phases,\brg}.} However, instead of continuing to negative
values on the far side, the volume is actually positive in the new
K\"ahler cone. In certain ${\cal N}=2$ heterotic/type IIA dual
pairs \hettwo, the singularity in the Calabi-Yau prepotential when
a curve in the type IIA geometry undergoes a flop (and an
effective gauge coupling becomes singular) is dual to a heterotic
strong coupling singularity. We describe one such example below.
It is important to stress that as expected on completely general
grounds, the effective $g^2$ remains positive everywhere in the
properly-interpreted type II moduli space.

The examples we have in mind, and their heterotic duals, are well
known. Our presentation of a specific example will closely follow
\louis, which mapped out in detail several heterotic/type II dual
pairs.

Let $X$ be the Calabi-Yau threefold which is an elliptic fibration
over $F_{1}$.  The prepotential for the K\"ahler moduli space of
$X$ is \louis:
\eqn\prepot{ {\cal{F}}_{II} = {4\over{3}}t_{1}^3 +
{3\over{2}}t_1^2 t_2 + {1\over{2}}t_1 t_2^2 + t_1^2 t_3 + t_1 t_2
t_3  } where $t_{i}$ are the K\"ahler moduli.  The volume of the
$(-1)$ curve is controlled by $t_{3}$. One can find a set of dual
heterotic variables $S,T,U$, which are related to the type II
variables by \eqn\transform{ t_1 = U,\qquad t_2 = T - U, \qquad
t_3 = S - {T\over{2}} - {U\over{2}} } In heterotic variables, the
prepotential reads \eqn\hetpot{ {\cal{F}}_{h}= STU +
{1\over{3}}U^3 } We know that the type II operation of shrinking
the $(-1)$ curve corresponds to strong gauge coupling in the
heterotic picture. This instructs us to identify
$S-{T\over{2}}-{U\over{2}}$ with the visible-sector gauge
coupling.\foot{To make contact with our earlier notation, $T$ and
$U$ are the two $T^{\alpha}$, and $\beta_{2}= {1\over{2}}$ for
$\alpha = 1,2$.}

Now, to study the effect of the strong gauge coupling, we flop the
curve corresponding to $t_{3}$.  The fields transform as
\eqn\flopgeo{(t_1,t_2,t_3) \rightarrow (t_1 + t_3,~ t_2 + t_3,~ -
t_3)} leading to the prepotential for $\tilde{X}$, the image of
$X$ under the flop.  It turns out that $\tilde{X}$ is not a $K3$
fibration, and furthermore it is not dual to a perturbative
heterotic model.

Given this linear implementation of the flop in type II variables,
we can apply this transformation to the heterotic variables (3.9).
This yields \eqn\flophet{ (U,~ T - U,~ S - {T\over{2}} -
{U\over{2}} ) \rightarrow (S + {U\over{2}}  - {T\over{2}},~ S
+ {T\over{2}}  - {3U\over{2}} ,~ {T\over{2}}  + {U\over{2}}  - S)}
The key result is that the visible-sector gauge coupling has
changed sign, \eqn\changesign{S- {T\over{2}}  - {U\over{2}}
\rightarrow - S +{T\over{2}}  + {U\over{2}}  } In this new
K\"ahler cone, the visible-sector coupling is sensible provided
$T+U > 2 S$, which is complementary to the initial restriction
$T+U < 2S$.

We have therefore seen that in this very simple example, the gauge
coupling function for the visible sector is sensible and positive
on both sides of the strong coupling transition. We expect this
result to hold in all of the cases of interest, simply from
macroscopic arguments about supersymmetric theories. It would be
interesting to generalize the simple illustration above to ${\cal
N}=1$ heterotic vacua by studying the dual geometric transitions
in F-theory compactifications on Calabi-Yau fourfolds.

\subsec{Fractional Invariants and Weak Coupling}

Let us now determine the conditions under which the stable vacuum
exists at modestly large values of $S$ and $T$.  Note that this
does not mean that all of the physics is weakly coupled, since as
we just discussed, we have undergone a strong coupling transition
in $E_8^{(1)}$! However, some other sectors of the theory may
remain perturbative at large $S$ and $T$, so it is still of
interest to know that stabilization at large $S$ and $T$ is
possible.

The goal is to arrange that the volume of the
Calabi-Yau is large in string units, while the string coupling is
small:\foot{For simplicity we now present the formulae for the
case of one K\"ahler modulus, the overall volume; the
generalization is straightforward.} \eqn\volume{ \left( S T\right)
^{ 1 \over 8} = e ^{\sigma}
> 1 ,} \eqn\dilaton{ \left({ T^3 \over S}\right)^{1 \over 2} =
e^{\phi} < 1.} Recall that $\phi$ is the ten-dimensional dilaton;
we denote the four-dimensional dilaton by $\varphi$.  Using the
relation \solutiona, we have \eqn\newvol{
{\left({3\over{\beta}}\right)} S^2 = e^{8 \sigma}} \eqn\newdil{
{\left({3\over{\beta}}\right)}^{3} S^2 = e^{\phi} } Clearly $\beta
> 3$ is a necessary condition for perturbative validity. It
follows from \betatwois\ that this condition can only be met if
the bundle $V_{2}$ is nontrivial; hence gaugino condensation in an
{\it{unbroken}} hidden-sector $E_8$ is not compatible with this
method of volume stabilization. To see explicitly that large
$\beta$ is possible within known constructions we refer to the
plots of \schimmrigk.

From the form of the solution \solutionb\ it is
clear that the values of $S$ and $T$ at the stable minimum
increase as the Chern-Simons invariant becomes smaller. We are
therefore interested in finding three-cycles admitting extremely
small Chern-Simons invariant.

Small values of the Chern-Simons invariant are distasteful but not
unattainable.  We saw in \S3 that it is possible to get a small
Chern-Simons invariant $h$ by working on a Calabi-Yau which has a
three-cycle $Q$ satisfying
$$\pi_1(Q) = \Z_{p}$$ for $p \gg 1$. The simplest example of this
is a Lens space. One way to generate even smaller $h$ is to take
$Q$ to be a general Seifert manifold $\Sigma(a_1,\ldots,a_n)$,
since the minimal value of $h$ would scale like
\eqn\hscales{h^{-1} \sim \prod_{i=1}^{n} a_i } With several $a_i$
one could then generate very small fractional fluxes.

\subsec{Summary of Requirements}

Let us briefly review the conditions on the Calabi-Yau $X$ and the
gauge bundles $V_{i}$ which ensure the existence of the
supersymmetric vacuum \solutionb\ with both dilaton and K\"ahler
moduli stabilized.  Conditions essential to the mechanism are
listed first, while those related to detailed model-building come
last.

\item{(1)}
In order to achieve a small value of the three-form flux, the
Calabi-Yau manifold $X$ must have a nontrivial fundamental group
and must admit gauge connections which have fractional
Chern-Simons invariant on a three-cycle $Q$ which is not torsion.
One of the bundles $V_{1}$,$V_{2}$ must then be chosen to have
such a gauge connection, i.e. suitable Wilson lines.  These
conditions are {\it{automatically}} met in a large class of
realistic string models.

\item{(2)}
For gaugino condensation to be possible in $H \subset E_8^{(2)}$,
the bundle $V_{2}$ must break $ E_8 \to H$ without introducing any
light charged matter, leaving a pure gauge group. For example, if
$H = E_7$ then there is no index theorem protecting charged matter
${\bf 56}$s, so we expect that this condition is generically
satisfied.  If instead $H = E_6$ the number of chiral generations
is $\vert {1\over{2}}\int_{X}c_{3}(V^{(2)}) \vert$. The bundle
$V_{2}$ should be chosen so that this vanishes.\foot{One could
imagine other possibilities in which charged matter in the hidden
sector generates a nonperturbative superpotential which can be
used for stabilization.  See e.g. \casas\ for a discussion of this
possibility in the context of racetrack models.}

\item{(3)}
In order to stabilize the overall volume we must choose bundles
for which the quantity $\beta_{2}$ defined in \betatwois\ is
nonzero. To stabilize multiple K\"ahler moduli we must take all of
the $\beta_{2}^{\alpha}$ to be nonzero. To ensure stabilization of
the volume above the string scale, we should also have $\beta_{2}
> 3$, with an analogous condition for the case of many moduli.

\item{(4)}
If the K\"ahler moduli are to be stabilized, the initial
configuration and the final stable minimum are on opposite sides
of a transition in which the visible sector becomes strongly
coupled. It follows that the visible-sector gauge theory can only
be properly understood in models where this strong coupling
transition can be followed in detail.  Better understanding of
this transition is a necessary prelude to the building of
realistic models.  Readers uncomfortable with the transition
are advised to set $\beta_{1}=\beta_{2}=0$, in which case one
is left with a no-scale model with fixed dilaton and an
unfixed volume modulus.

\item{ (5)}
Further constraints will be necessary to obtain realistic
low-energy physics.  For example, $V_{1}$ should contain
appropriate Wilson lines which break the visible-sector GUT to the
Standard Model gauge group.  (It is sometimes possible to arrange
that these same Wilson lines also provide the fractional
Chern-Simons invariant.)  The vacua we have constructed have
negative cosmological constant, with an energy density not far
below the string scale.  This must certainly be modified to lead
to a sensible cosmological model!
Finally, if we wish to stabilize at very weak coupling then the
fundamental group of the Calabi-Yau must be unusually large.

Clearly, the greatest obstacle to calculability in this scenario
is the strong coupling transition in the observable sector. It is
conceivable that one could avoid this difficulty by combining
fractional Chern-Simons invariants and gaugino condensation with a
non-K\"ahler compactification geometry, for in this case the
volume modulus can be stabilized at tree level.  However, for the
bulk of our analysis, the only real assumption we have made is
that the unknown physics of the visible sector does not modify the
superpotential. This seems believable provided that the low-energy
${\cal N}=1$ gauge theory which emerges from $E_8^{(1)}$ is not
one which dynamically generates a superpotential.


\newsec{Duality to Type IIA and M-theory}

The models studied in this paper are related by various dualities
to a particular class of ${\cal N}=1$ compactifications of
M-theory and Type IIA string theory.  These models have recently
received some attention due in part to phenomenological
applications, see {e.g.}
\refs{\AW,\AcharyaW,\Wdeconstr,\FW,\KlebanovW,\Bobby,\CveticSU,\CveticLW}.
After appropriate duality transformations our mechanism for moduli
stabilization can be applied to these models as well. In this
section we briefly discuss various aspects of these dualities, as
well as their implications.

\subsec{Heterotic/Type IIA Duality}

Our considerations have thus far been limited to the $E_8 \times
E_8$ heterotic string, but the discussion can be repeated almost
verbatim for the $Spin(32)/{\Z_{2}}$ heterotic string compactified
on a Calabi-Yau manifold.  The latter theory is related to an
${\cal N}=1$ compactification of type IIA string theory by the
following chain of dualities:
\eqn\hetsodual{ Spin(32)/{\Z_{2}} {\rm ~Het~} \buildrel S \over
\longleftrightarrow {\rm ~Type~I~} \buildrel \simeq \over
\longleftrightarrow {\rm ~Type~IIB~}/\Omega \buildrel T \over
\longleftrightarrow {\rm ~Type~IIA~}/(\Omega \cdot {\cal I}) }
Let us now explain each step in this duality in more detail and,
in particular, find the relation between the parameters and the
coupling constants. The first relation is the standard strong-weak
coupling duality between the $Spin(32)/{\Z_{2}}$ heterotic string
theory and type I string theory. The effective supergravity action
in the latter theory is similar to the heterotic supergravity
action, with the type I and heterotic variables related by
%
%
%
\eqn\ihetphi{\phi_I = - \phi_H}
\eqn\ihetg{g^{I}_{MN} = g^{H}_{MN} e^{- \phi_H}}
At the next step in the chain of dualities \hetsodual\ we identify
type I string theory with an orientifold of type IIB closed string
theory, where $\Omega$ denotes the world-sheet parity symmetry.
The parameters and the coupling constants in the supergravity
action do not change under this identification, although some
terms acquire a different interpretation. In particular, in the
type IIB theory the gauge degrees of freedom arise as open string
states on the world-volume of 32 space-filling D9-branes. Thus,
the Wilson lines of the original heterotic string theory become
Wilson lines on D9-branes, and the ten-dimensional gauge coupling
is simply
\eqn\iibgym{{g_{10}^2 \over \alpha'^3} = e^{\phi_I} =
e^{\phi_{IIB} }}
From \ihetphi\ and \iibgym\ we find \eqn\hetB{\phi_{IIB} = \phi_I
= - \phi_H}
The last step in \hetsodual\ is the T-duality --- mirror symmetry,
to be more precise --- between type IIB string theory on a
Calabi-Yau manifold $X$ and type IIA theory on the mirror manifold
$\tilde X$. Strictly speaking, the dual background is an
orientifold of $\tilde X$, where the involution changes the
orientation of the $T^3$ fibers. Under T-duality, the
space-filling D9-branes transform into D6-branes wrapped over the
base, $Q$, of the special Lagrangian torus fibration \SYZ. The
parameters of the resulting type IIA background can be obtained
from the usual T-duality rules:
\eqn\iiahphi{\phi_{IIA} = \phi_{IIB} - \log \left( { V_{X}^{IIB}
\over {V^{IIB}_Q\alpha'^{3/2}} } \right) = {1 \over 2} \phi_H -
\log \left( { V_{X}^{H} \over {V^{H}_Q\alpha'^{3/2}} } \right) }
Here $V_{X}$ and $V_Q$ denote, respectively, the volume of the
Calabi-Yau space $X$ and the volume of the base three-manifold $Q$
in the string theory given by the superscript.

To summarize, after a chain of dualities \hetsodual\ we found that
our heterotic string models are dual to IIA string theory on a
mirror Calabi-Yau manifold $\tilde X$, with D6-branes wrapped over
the special Lagrangian three-cycle $Q$. This is precisely the
configuration studied in
\refs{\AcharyaW,\Wdeconstr,\FW,\KlebanovW}. In these papers, $Q$
is usually taken to be a Lens space, $Q = {\bf S}^3 / \Z_p$, and
the Calabi-Yau manifold $\tilde X$ is usually assumed to be
non-compact. If $\tilde X$ is compact, as described above, then
the presence of orientifold 6-planes is crucial to cancel the
D6-brane charge.

Observe that on the D6-brane world-volume there is a topological
coupling between the gauge field, $F=dA + A \wedge A$, and the
Ramond-Ramond tensor fields $C = C_1 + C_3 + \ldots$,
\eqn\wzterms{ \tr \int_{{\bf R}^4 \times Q}  C \wedge e^{F} }
Among other terms, this expression contains a coupling
\eqn\ffour{ CS(A,Q) \int_{{\bf R}^4 }  G }
which we obtained by expanding \wzterms\ and integrating by parts.
It follows that D6-branes wrapped over $Q$ with a non-zero value
of the Chern-Simons invariant act as an effective source for the
Ramond-Ramond four-form field strength in the four uncompactified
directions.

\bigskip {\noindent {\it{ Comments on Proton Decay }}}

Using the chain of dualities \hetsodual\ we have now related our
setup to compactifications of type IIA string theory, where the
GUT gauge theory is realized on the world-volume of D6-branes
wrapped over a compact three-manifold $Q$.  Similar configurations
have been discussed in a recent work of Klebanov and Witten
\KlebanovW\ (see also \CveticP),
where it was shown that the proton decay rate from
dimension six operators is given by\foot{For simplicity, we omit
numerical factors of order one.}
\eqn\kwrate{ A_{IIA} \sim {g_{YM}^{4/3} L(Q)^{2/3} e^{{1 \over 3}
\phi_{IIA}} \over M_{GUT}^2 } }
where $g_{YM}$ is the GUT gauge coupling, and $M_{GUT}$ is the
unification scale. This scale is determined by the size of the
three-manifold $Q$,
\eqn\iiamgut{ M_{GUT} = \left( {L(Q) \over V_Q} \right)^{1/3} }
where the extra factor $L(Q)$ accounts for the one-loop threshold
corrections from Kaluza-Klein harmonics on $Q$
\refs{\FW,\KlebanovW}. Specifically, $L(Q)$ is a topological
invariant of $Q$, known as the Ray-Singer torsion.

Let us now compute the proton decay rate in our heterotic models.
In contrast to the result of \KlebanovW, we expect in our case the
conventional amplitude
\eqn\usualrate{ A_{h} \sim {g_{YM}^2\over{M_{GUT}^2}} \sim \alpha'
e^{{\phi\over{2}}+2\sigma} } where the unification scale and the
gauge coupling are given by \gymviav\ and \kkis, respectively.
By tracing the chain of dualities
\hetsodual\ in reverse, being careful to include the constant
rescaling of the Einstein-frame metric mentioned in \S2, one can
verify that \kwrate\ and \usualrate\ differ by the factor
$\alpha_{GUT}^{1/3} e^{-{1\over{3}}\phi_{IIA}}$, which exhibits
the anomalous scaling with $\alpha_{GUT}$ explained in \KlebanovW.


\subsec{Lift to M-theory}

Now let us consider the M-theory lift of the type IIA
configuration considered above. Since D6-branes wrapped over a
special Lagrangian submanifold $Q \subset \tilde X$ preserve
${\cal N}=1$ supersymmetry in four dimensions, their lift to
M-theory must be described by a seven-dimensional manifold,
$X_{G_2}$, with $G_2$ holonomy. Topologically, $X_{G_2}$ can be
viewed as a $K3$ fibration over $Q$ \GYZ,
\eqn\kthreebdle{
\matrix{ K3 & \rightarrow & X_{G_2} \cr  & & \downarrow \cr && Q} }
such that each $K3$ fiber has an ADE singularity, which
corresponds to the type of the gauge group on the D6-branes. For
example, $SU(5)$ gauge theory would lift to a $G_2$-manifold with
$A_4$ singularities in the fiber. The dual M-theory geometry
\kthreebdle\ can be obtained directly from the heterotic string
theory on a Calabi-Yau manifold $X$ by using the familiar duality
between M-theory on $K3$ and heterotic string theory on $T^3$.
Applying this duality to each fiber in the special Lagrangian
torus fibration, $X \to Q$, we end up with M-theory on a
seven-manifold $X_{G_2}$ with $G_2$ holonomy and topology
\kthreebdle. Various aspects of M-theory on $G_2$-manifolds of
this kind have been studied in
\refs{\AW,\AcharyaW,\Oz,\Friedmann,\Wdeconstr,\FW}.

Now let us consider a D6-brane configuration with non-trivial
gauge fields characterized by $CS(A,Q) \ne 0$. According to
\ffour, such gauge fields act as a source (localized on the
three-cycle $Q$) for the space-time component of the four-form
flux, $G_{0123}$. In the effective four-dimensional field theory,
this means there is a non-zero superpotential induced by
$CS(A,Q)$. In M-theory, the relevant interaction term \ffour\
appears due to anomaly inflow at the location of ADE singularities
\Wgtwo, while the effective superpotential is generated by
topologically non-trivial gauge fields supported at the
singularities \Bobby.

The models studied in this paper have real values of the
Chern-Simons invariant $CS(A,Q)$.  However, Acharya has argued
\Bobby\ that, in a more general setting, the superpotential
induced by gauge fields should be given by a {\it complex}
Chern-Simons invariant.  A deeper understanding of the connection
between these ideas would be quite interesting.


\newsec{Domain Walls}

In order to obtain an expression for the effective superpotential
of an ${\cal N}=1$ supersymmetric gauge theory, it is often useful
to study the spectrum of BPS domain walls.  Moreover, in a theory
with gaugino condensation, the domain walls provide information
about the breaking of chiral symmetry and about other phenomena of
interest.

With this motivation in mind, let us consider domain walls in our
models\foot{For a related discussion see also
\refs{\triples,\sethi}.}, where different vacua are characterized
by the values of the Chern-Simons functional, $CS(A,Q)$. Hence,
the BPS domain walls are represented by self-dual field
configurations (instantons) supported on $Q \times {\bf R}$, where
${\bf R}$ represents a spatial direction orthogonal to the domain
wall. Since $CS(A,Q)$ takes fractional values, such instantons
carry fractional charge,
\eqn\instnumber{ c_2 = - {1 \over 8 \pi^2}  \int {\rm{tr}}\left(F
\wedge F\right) = CS(A,Q)\vert_{- \infty} - CS(A,Q)\vert_{+
\infty}}
The instanton action is given by $\int_{Q \times {\bf R}}\tr
\left(F \wedge* F\right)$, which, using the self-duality of the
gauge field $F$, can be written as
$$
 \int_{Q \times {\bf R}} \tr \left(F \wedge F\right)
$$
Furthermore, using \instnumber\ one can rewrite the instanton action
as the difference of the values of the Chern-Simons functional,
$\Delta CS(A,Q)$.
Comparing this formula with the standard expression for the tension
of a domain wall in ${\cal N}=1$ supersymmetric theory,
$T= \vert \Delta W \vert$,
we come to our previous result \wfis\ for the effective
superpotential induced by non-trivial gauge fields \refs{\GVW,\calibr}
\eqn\wcsis{ W_{flux} = \int_Q \Omega_3 (A) }

Now let us consider the degeneracy of domain walls interpolating
between two vacua with fractional Chern-Simons functional,
$CS(A,Q)$, for some three-cycle $Q \subset X$. At least in the
classical theory, the BPS domain walls come in continuous
families. Specifically, the moduli space of domain walls with
fractional charge $c_2$ is isomorphic to the moduli space of
charge-$c_2$ instantons on $Q \times {\bf R}$,
\eqn\mspace{ {\cal M} \left(Q \times {\bf R} ; c_2 \right) }
Without loss of generality, we can study $SU(2)$ instantons
and, for concreteness, take $Q$ to be a Lens space,
$$
Q = {\bf S}^3 / \Z_p
$$
Then, according to \lenscs, the Chern-Simons functional on $Q$ can
take the fractional values:
\eqn\lenscsa{ CS(A,Q) = - {m^2 \over 4p} - {\lambda^2 \over 2}}
Here we follow the notations of \Austin, introduced at the end of
\S3, where $m$ is an integer defined modulo $2p$.

Consider an instanton on $Q \times {\bf R}$ which interpolates
between different values of the Chern-Simons invariant, $CS(A,Q)$.
According to \instnumber\ and \lenscsa, such an instanton connects
two states characterized by different rotation numbers $m$ and
$m'=m$ mod 2, and carries a fractional instanton charge, $c_2 =
k/p$. Put differently, it is described by a triplet of integers,
$(k,m,m')$. Following \Austin, let us express $(m,m') \sim
(a-b,a+b)$ in terms of $a$ and $b$, such that
\eqn\twoeqs{\eqalign{ a & = (m' + m)/2 \quad\quad {\rm mod~} p \cr
b & = (m' - m)/2 \quad\quad {\rm mod~} p }}

Using the above expression \lenscsa\ for the value of
the Chern-Simons functional, we find the corresponding
instanton number:
$$
\eqalign{
c_2 & = CS(A,Q)\vert_{- \infty} - CS(A,Q)\vert_{+ \infty}
= \cr
& = - {(a-b)^2 \over 4p} + {(a+b)^2 \over 4p} = {ab \over p} }
$$
Therefore, we have
\eqn\kviaab{ k = ab \quad\quad {\rm mod~} p }

Now we are in a position to describe the moduli space,
${\cal M}$, of instantons on $Q \times {\bf R}$
that interpolate between gauge connections
with rotation numbers $m = a-b$ and $m'=a+b$.
Since instanton configurations always have
a modulus that represents their position in ${\bf R}$,
it makes sense to divide by translations
and consider the reduced moduli space,
$$
{\cal M}' = {\cal M}/{\bf R}
$$
Using index theorems one can compute the virtual dimension of the
reduced moduli space \Austin,
\eqn\mdim{
{\rm Dim} ({\cal M}') = {8k \over p} - 4 + n
+ {2 \over p} \sum_{j=1}^{p-1} \cot^2 {\pi j \over p}
\left( \sin^2 {\pi j m \over p} - \sin^2 {\pi j m' \over p} \right)
}
where $n \in \{ 0,1,2 \}$ is the number of $m$, $m' \ne 0,p$. It
turns out that this virtual dimension is always even. In order to
illustrate this general formula, in the table below we list the
dimensions of the moduli spaces of fractional charge instantons on
${\bf S}^3/ \Z_5 \times {\bf R}$. In terms of $a$ and $b$,
$m=a-b$, $m'=a+b$, and the instanton number $k=ab$ mod 5.

\vskip 0.8cm
\vbox{
\centerline{\vbox{
\hbox{\vbox{\offinterlineskip
\def\tablespace{height7pt&\omit&&\omit&&\omit&&\omit&&\omit&&\omit&\cr}
\def\tablerule{\tablespace\noalign{\hrule}\tablespace}
\hrule\halign{&\vrule#&\strut\hskip0.2cm\hfill #\hfill\hskip0.2cm\cr
\tablespace
& $a \backslash b$   && 0 && 1 && 2 && 3 && 4 &\cr
\tablerule
& 0 && -- && --  && --  && --  && --  &\cr
\tablerule
& 1 && --  && 0  && 2  && 2  && 2  &\cr
\tablerule
& 2 && --  && 2  && 4  && 6  && 10  &\cr
\tablerule
& 3 && --  && 2  && 6  && 12  && 18  &\cr
\tablerule
& 4 && --  && 2  && 10  && 18 && 24  &\cr
\tablespace
}\hrule}}}}
\centerline{
\hbox{{\bf Table 1:} ~~Dim$({\cal M}')$ {\it for the Lens space,
$Q = {\bf S}^3/\Z_5$.}}}
}
\vskip 0.5cm

The dimension of the moduli space tends to grow with the instanton
number, $k=ab$. For low values of the dimension, one can describe
${\cal M}'$ rather explicitly using general topological properties
\Austin (see also \refs{\FS,\FL}). When Dim$({\cal M}')=0$, the
reduced moduli space must be just a point. In this case, we have
only one domain wall interpolating between two vacua. Furthermore,
the Euler number of ${\cal M}'$ is given by the number of
solutions $(a,b)$ to the equations \twoeqs, such that $ab=k$. In
particular, this implies that
\eqn\positiveeuler{ \chi ({\cal M}') \ge 0 }
Hence, when Dim$({\cal M}')=2$, the reduced moduli space must
be of the form,
$$
{\cal M}' = {\bf S}^2 \setminus F
$$
where $F$ is a set of 0, 1, or 2 points.

For example, let us take $p=5$, $a=2$, and $b=1$. This implies
$k=2$, $m=1$, and $m'=3$. Then, from Table 1 we find that ${\cal
M}'$ must be of real dimension 2, and by looking at the Euler
number $\chi ({\cal M}')=2$ one concludes that in this example the
moduli space is simply a two-sphere,
$$
{\cal M}' = {\bf S}^2
$$
Since this space is compact, we expect that the degeneracy of
domain walls of charge $c_2 = 2/5$ interpolating between vacua
with $m=1$ and $m'=3$ is given by the cohomology of ${\cal M}'$.
Therefore, in this example we find
$$
\# ({\rm ~domain~walls~}) = 2
$$

The above results suggest the following conjecture for the degeneracy
of domain walls with small fractional charge, $c_2 = k/p$,
\eqn\instconj{\# ({\rm ~domain~walls~}) = \cases{1 & if $k=1$ \cr2
& if $k=2$  }}
%
In other words, we expect that there is always only one domain wall
of the minimal fractional charge, whereas the degeneracy of
domain walls with twice the minimal charge is equal to 2.
It would be interesting to pursue this analysis further.

\newsec{Discussion}

We have argued that it is possible to stabilize the complex
structure moduli, K\"ahler moduli, and dilaton of heterotic
Calabi-Yau compactifications.  Our ingredients are hidden-sector
gaugino condensation combined with a flux-generated superpotential
arising from a flat connection with fractional Chern-Simons
invariant. For the non-K\"ahler compactifications of
\refs{\nonkahler,\bbdp} our result looks even more promising,
since there the volume is stabilized at tree level, and the only
concern is the dilaton.

One omission from our list of stabilized moduli is the vector
bundle moduli.  Following the analysis in \edold, it seems
likely that the very existence of bundle moduli is not generic.
Massless modes arising from the moduli of a vector bundle $V$ are
associated with elements of the group $H^{1}(X, End(V))$.
Typically there is no index theorem which allows one to argue that
this group should be nontrivial. Even if the group were
nontrivial, a generic infinitesimal deformation of the vector
bundle is obstructed at some finite order and so does not
constitute a modulus.\foot{Nevertheless, simple bundles
constructed by mere humans often have moduli.  In many such simple
cases, even nonperturbative sigma model effects do not suffice to
lift them \wsinst.  Examples of superpotentials arising for the
bundle moduli associated with small instantons in heterotic
M-theory are described in e.g. \ovrut.}

In addition to the omission of a detailed discussion of bundle
moduli, we have used standard approximations in describing the
hidden sector gaugino condensation.  For instance, in real string
models, the hidden sector would have massive fields charged under
the hidden $E_8$.  This would lead to corrections to the form of
the superpotential used here, which presumably arise as more
highly damped exponentials in $S$. While for reasonable values of
$Re(S)$ this should not be a large correction, it would be nice to
have exact results.  These are not yet available for
${\cal N}=1$ supersymmetric compactifications of heterotic strings.

The solutions we have constructed are supersymmetric AdS vacua. It
is natural to ask whether one can add a source of
supersymmetry-breaking energy which lifts these models to de
Sitter vacua, along the lines of \kklt. In fact, there are
significant similarities between the type IIB constructions of
\kklt\ and the heterotic models discussed here. As noted in \S4,
the superpotential in each case consists of a small, constant term
from flux and an exponential term from nonperturbative gauge
dynamics. To continue this analogy and
include supersymmetry breaking, one would have to
introduce the heterotic dual of the anti-D3-brane introduced
in \kklt.
In heterotic M-theory this would correspond to a non-supersymmetric
wrapped M5-brane.
To achieve control over the construction, one would need to introduce
such an object in a heterotic background with significant warping.

Burgess, Kallosh, and Quevedo \bkq\ have recently proposed that a
Fayet-Iliopoulos D-term potential could serve as another useful
source of energy for uplifting heterotic models. The stable AdS
vacua we have discussed would appear to be a suitable setting for
such a mechanism, but we leave the construction of explicit models
as a subject for future exploration.  Again, one would have to
arrange for a suitably small D-term to justify the analysis.

The present proposal for manufacturing vacua without moduli, combined
with the
constructions in \refs{\kklt,\Bobby,\bkq,\Eva,\otherc}, is a small step
towards filling out our picture of the ``discretuum'' \BP\ of
string/M-theory vacua.  This is the full space of vacua of string
theory, including all of the possibilities for the background
fluxes, wrapped branes, and other discrete data. Interesting
general aspects of this landscape of string theory vacua have
recently been discussed in e.g
\refs{\douglas,\susskind,\AcharyaYuk}, while statistical arguments
relying on the existence of the discretuum have been used in e.g.
\refs{\BP,\wilczek,\kklt} in tuning the cosmological constant.

Although this is a bit far from the concrete goal of our paper, it
is worth discussing how this discretuum may be expected to arise
in the heterotic theory.  In type II theories, as in M-theory, the
discretuum is populated by vacua with various quantized values of
the RR and NS fluxes, and with different wrapped branes,
consistent with the tadpole conditions arising from the Gauss' law
constraints on the various p-form field strengths.  In the
heterotic theory, there are a few quantum numbers which contribute
to the large number of vacua.  In addition to the large number of
choices of vector bundles on a fixed manifold (characterized by
the topological numbers $c_{2}(V_i)$, $c_3(V_i)$, for instance),
there are also background NS fluxes. Finally, there is the
possibility of non-K\"ahlerity, which is roughly dual to the
possibilities of different fluxes in type II theories \nonkahler.

As described at length in \douglas, to get a good handle on this large set
of possibilities, it will probably be necessary to find auxiliary
ensembles which accurately model the space of vacua.
We have little to say about this at present but leave it as an ambitious
goal for future research.

\medskip
\bigbreak \centerline{\bf{Acknowledgements}}

We would like to thank B. Acharya, N. Arkani-Hamed, P. Aspinwall,
M. Becker, K. Dasgupta, M. Dine, M. Douglas, R. Kallosh, A.
Krause, E. Silverstein, S. Thomas, S. Trivedi and E. Witten for
interesting discussions on related subjects.  This work was
conducted during the period S.G. served as a Clay Mathematics
Institute Long-Term Prize Fellow. S.G. is also supported in part
by RFBR grant 01-01-00549 and RFBR grant for Young Scientists
02-01-06322. The work of S.K. is supported in part by a David and
Lucile Packard Foundation Fellowship for Science and Engineering,
National Science Foundation grant PHY-0097915, and the DOE under
contract DE-AC03-76SF00515.  The work of L.M. is supported by a
National Science Foundation Graduate Research Fellowship.

\listrefs
\end